\documentstyle[11pt,aaspp4,flushrt,natbib209]{article}
\slugcomment{\it Accepted to the Astrophysical Journal (April 24, 2012)}
\lefthead{Stern et al.}
\righthead{WISE AGN in COSMOS}

\def\ie{{i.e.}}
\def\eg{{e.g.}}
\def\etal{{et al.}}
\def\wise{{\it WISE}}
\def\chandra{{\it Chandra}}
\def\xmm{{\it XMM-Newton}}
\def\spitzer{{\it Spitzer}}

\def\deg{\ifmmode {^{\circ}}\else {$^\circ$}\fi}
\def\kms{\ifmmode {\rm\,km\,s^{-1}}\else
    ${\rm\,km\,s^{-1}}$\fi}
\def\ergcm2s{\ifmmode {\rm\,ergs\,cm^{-2}\,s^{-1}}\else
    ${\rm\,ergs\,cm^{-2}\,s^{-1}}$\fi}
\def\ergAcm2s{\ifmmode {\rm\,ergs\,cm^{-2}\,s^{-1}\,\AA^{-1}}\else
    ${\rm\,ergs\,cm^{-2}\,s^{-1}\,\AA^{-1}}$\fi}
\def\ergs{\ifmmode {\rm\,ergs\,s^{-1}}\else
    ${\rm\,ergs\,s^{-1}}$\fi}
\def\kmsMpc{\ifmmode {\rm\,km\,s^{-1}\,Mpc^{-1}}\else
    ${\rm\,km\,s^{-1}\,Mpc^{-1}}$\fi}

\def\spose#1{\hbox to 0pt{#1\hss}}
\def\simlt{\mathrel{\spose{\lower 3pt\hbox{$\mathchar"218$}}
     \raise 2.0pt\hbox{$\mathchar"13C$}}}
\def\simgt{\mathrel{\spose{\lower 3pt\hbox{$\mathchar"218$}}
     \raise 2.0pt\hbox{$\mathchar"13E$}}}

\def\plotfiddle#1#2#3#4#5#6#7{\centering \leavevmode
\vbox to#2{\rule{0pt}{#2}}
\includegraphics{#1}}

\begin{document}

\title{Mid-Infrared Selection of AGN with the Wide-Field Infrared Survey Explorer. I.  Characterizing WISE-Selected AGN in COSMOS}

\author{Daniel Stern\altaffilmark{1},
Roberto J. Assef\altaffilmark{1,2},
Dominic J. Benford\altaffilmark{3},
Andrew Blain\altaffilmark{4},
Roc Cutri\altaffilmark{5},
Arjun Dey\altaffilmark{6},
Peter Eisenhardt\altaffilmark{1},
Roger L. Griffith\altaffilmark{5},
T.H. Jarrett\altaffilmark{5},
Sean Lake\altaffilmark{7},
Frank Masci\altaffilmark{5},
Sara Petty\altaffilmark{7},
S.A. Stanford\altaffilmark{8,9},
Chao-Wei Tsai\altaffilmark{5},
E.L. Wright\altaffilmark{7},
Lin Yan\altaffilmark{5},
Fiona Harrison\altaffilmark{10}
\& Kristin Madsen\altaffilmark{10}}

\altaffiltext{1}{Jet Propulsion Laboratory, California Institute of
Technology, 4800 Oak Grove Drive, Mail Stop 169-221, Pasadena, CA
91109 [e-mail: {\tt daniel.k.stern@jpl.nasa.gov}]}

\altaffiltext{2}{NASA Postdoctoral Program Fellow}

\altaffiltext{3}{NASA Goddard Space Flight Center, Code 665,
Greenbelt, MD 20771}

\altaffiltext{4}{Department of Physics and Astronomy, University
of Leicester, LE1 7RH Leicester, UK}

\altaffiltext{5}{Infrared Processing and Analysis Center, California
Institute of Technology, Pasadena, CA 91125}

\altaffiltext{6}{National Optical Astronomical Observatory, 950 N.
Cherry Ave., Tucson, AZ 85719}

\altaffiltext{7}{Physics and Astronomy Department, University of
California, Los Angeles, CA 90095}

\altaffiltext{8}{Department of Physics, University of California,
One Shields Avenue, Davis, CA 95616}

\altaffiltext{9}{Institute of Geophysics and Planetary Physics,
Lawrence Livermore National Laboratory, Livermore, CA 94550}

\altaffiltext{10}{Space Radiation Laboratory, California Institute
of Technology, Pasadena, CA 91125}

\begin{abstract} 

The {\it Wide-field Infrared Survey Explorer (WISE)} is an extremely
capable and efficient black hole finder.  We present a simple
mid-infrared color criterion, $W1 - W2 \geq 0.8$ (\ie, [3.4]$-$[4.6]
$\geq 0.8$, Vega), which identifies $61.9 \pm 5.4$ AGN candidates
per deg$^2$ to a depth of $W2 \sim 15.0$.  This implies a much
larger census of luminous AGN than found by typical wide-area
surveys, attributable to the fact that mid-infrared selection
identifies both unobscured (type 1) and obscured (type 2) AGN.
Optical and soft X-ray surveys alone are highly biased towards only
unobscured AGN, while this simple \wise\, selection likely identifies
even heavily obscured, Compton-thick AGN.  Using deep, public data
in the COSMOS field, we explore the properties of \wise-selected
AGN candidates.  At the mid-infrared depth considered, 160 $\mu$Jy
at 4.6 $\mu$m, this simple criterion identifies 78\% of {\it Spitzer}
mid-infrared AGN candidates according to the criteria of Stern et
al. (2005) and the reliability is 95\%.  We explore the demographics,
multiwavelength properties and redshift distribution of \wise-selected
AGN candidates in the COSMOS field.

\end{abstract}


\section{Introduction}

Most surveys for AGN are severely biased towards unobscured (type~1)
AGN.   Nuclear emission in such sources dominates over host galaxy
light at most wavelengths, making type~1 AGN both more readily
identifiable and easier to follow-up spectroscopically.  However,
models predict a large population of obscured (type~2) AGN,
outnumbering type~1 AGN by a factor of $\sim 3$ \citep[\eg,][]{Comastri:95,
Treister:04, Ballantyne:11}.  Determining the ratio of unobscured
to obscured AGN as a function of luminosity and redshift has direct
implications for the growth history of supermassive black holes in
galactic centers, as well as for the origin of the cosmic X-ray
background (and, at a $\sim 10$\%\ level, optical and infrared
backgrounds).  Furthermore, recent theoretical work suggests that
AGN feedback plays a dominant role in establishing the present-day
appearances of galaxies \citep[\eg,][]{Silk:98, Hopkins:08}.  With the
dominant population of obscured AGN severely underrepresented by
current studies, however, a full understanding of the interplay
between AGN feedback and galaxy formation is hampered.

The most promising photometric techniques for identifying luminous
type~2 AGN are radio selection, hard X-ray selection and mid-infrared
selection.  However, only $\sim 10\%$ of AGN are radio-loud
\citep[\eg,][]{Stern:00a} and the current generation of hard X-ray
satellites have limited sensitivity.  Specifically, recent surveys
with {\it Swift} and {\it INTEGRAL} have only identified a few dozen
heavily obscured (\eg, Compton-thick) AGN, all at very low redshift,
$z \approx 0$ \citep[\eg,][]{Bassani:99, Vignali:02, Ajello:08,
Tueller:08, Burlon:11}.  The {\it Nuclear Spectroscopic Telescope
Array (NuSTAR)} \citep{Harrison:10}, scheduled for launch in early
2012, will improve that hard X-ray ($\sim 30$~keV) sensitivity by
a factor of $\sim 200$, but with a field of view comparable to {\it
Chandra}/ACIS, {\it NuSTAR} will only undertake a limited number
of extragalactic surveys, unlikely to cover more than a few square
degrees of sky.  With mid-infrared sensitivities several orders of
magnitude greater than the {\it Infrared Astronomical Satellite
(IRAS)}, the {\it Wide-field Infrared Survey Explorer (WISE)}
\citep{Wright:10} promises the first sensitive full-sky survey for
both type~1 and type~2 luminous AGN.

\wise\, launched on UT 2009 December 14 and completed its first
survey of the entire sky on UT 2010 July 17, obtaining a minimum
coverage of five exposures per sky position over 95\% of the sky 
in four passbands, 3.4, 4.6, 12 and 22 $\mu$m ($W1,
W2, W3$ and $W4$).  The all-sky data release occurred on 2012 March
14, releasing all data taken during the \wise\, full cryogenic
mission phase\footnote{See {\tt
http://wise2.ipac.caltech.edu/docs/release/allsky/}.}.  The median depth-of-coverage
is 15.6 exposures per sky position for $W1$ and $W2$, and 14.8 exposures per
sky position for $W3$ and $W4$.   Accounting
for source confusion, the estimated sensitivities are 0.068, 0.098, 0.86
and 5.4 mJy ($5 \sigma$), respectively \citep{Wright:10}.  The corresponding Vega
magnitude limits are 16.83, 15.60, 11.32, and 8.0, respectively.  The depth
increases with ecliptic latitude, reaching more than five times
greater sensitivity near the ecliptic poles \citep{Jarrett:11}.  In
order of increasing wavelength, the imaging resolution (FWHM) is
6\farcs1, 6\farcs4, 6\farcs5 and 12\farcs0 for the four bands.

Similar to the UV-excess method of identifying quasars
\citep[\eg,][]{Schmidt:83}, mid-infrared selection of AGN relies
on distinguishing the approximately power-law AGN spectrum from the
black body stellar spectrum of galaxies which peaks at rest-frame
$1.6 \mu$m.  Mid-infrared data easily separate AGN from stars and
galaxies, with the added benefit that mid-infrared selection is
less susceptible to dust extinction and is sensitive to the highest
redshift sources.  Courtesy the unprecedented sensitivity and mapping
efficiency of the Infrared Array Camera \citep[IRAC;][]{Fazio:04a}
onboard the {\it Spitzer Space Telescope}, the past few years have
seen an explosion of research using mid-infrared observations to
find and study (obscured) AGN at high redshift \citep[\eg,][]{Lacy:04,
Lacy:07, Stern:05b, AlonsoHerrero:06, Barmby:06, MartinezSansigre:06,
MartinezSansigre:07, Donley:07, Donley:08, Donley:12, Dey:08,
Fiore:08, Fiore:09, Hatziminaoglou:08, Rigopoulou:09, Seymour:07,
DeBreuck:10, Eckart:10, Park:10}.  IRAC identification of AGN typically required
all four passbands of that instrument, with data out to $8 \mu$m,
to differentiate AGN from high-redshift ($z \simgt 1.3$) massive
galaxies.  This is because distant, massive galaxies have red
observed colors from 3 to 5 $\mu$m \citep[\eg,][]{Lacy:04, Stern:05b,
Donley:07, Donley:12, Hickox:09, Galametz:12} and even very shallow
($< 90$~sec) IRAC pointings easily reach well below the characteristic
brightness of early-type galaxies out to $z \sim 2$, $m^*_{4.5}
\approx 16.7$ \citep[\eg,][]{Mancone:10}.  Courtesy of its shallow
observations, \wise\, suffers less pronouncedly from such contamination
and therefore is able to robustly identify AGN with just the two
bluest, most sensitive channels.

\begin{figure}[t!]
\plotfiddle{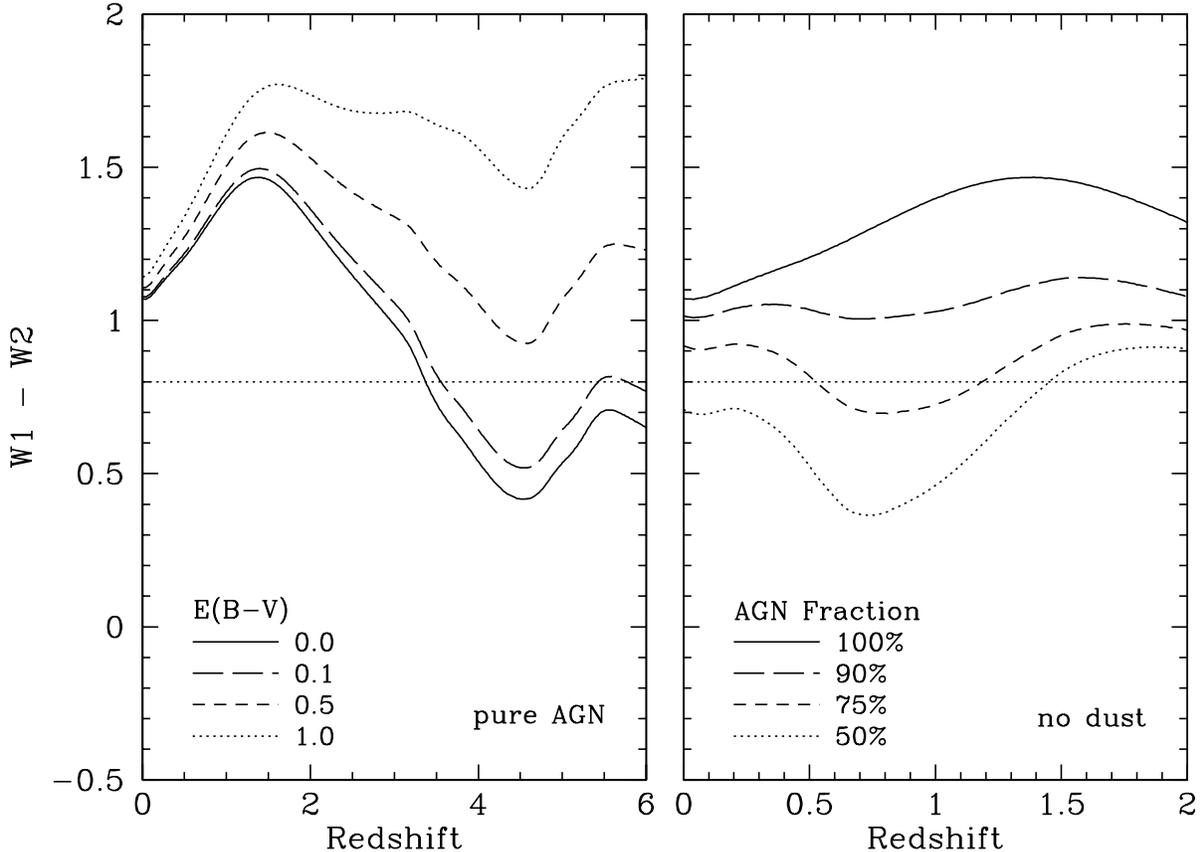}{4.5in}{-90}{60}{60}{-240}{360}
\caption{Model colors of AGN as a function of redshift using the
templates of Assef et al. (2010; Vega magnitudes).  The left panel
shows a pure AGN template with increasing amounts of dust extinction.
The right panel shows an unextincted AGN diluted by increasing
amounts of host galaxy light, where the host is the early-type (E)
template from Assef et al. (2010); changing galaxy template has
minimal effect.   A simple color criterion of $W1 - W2 \geq 0.8$
identifies pure AGN out to $z \sim 3$ and extincted pure AGN out
to higher redshifts.  For unextincted AGN, sources are no longer
selected as the host galaxy becomes an increasing fraction of the
bolometric luminosity.
\label{fig:assef}}
\end{figure}

\begin{figure}[t!]
\plotfiddle{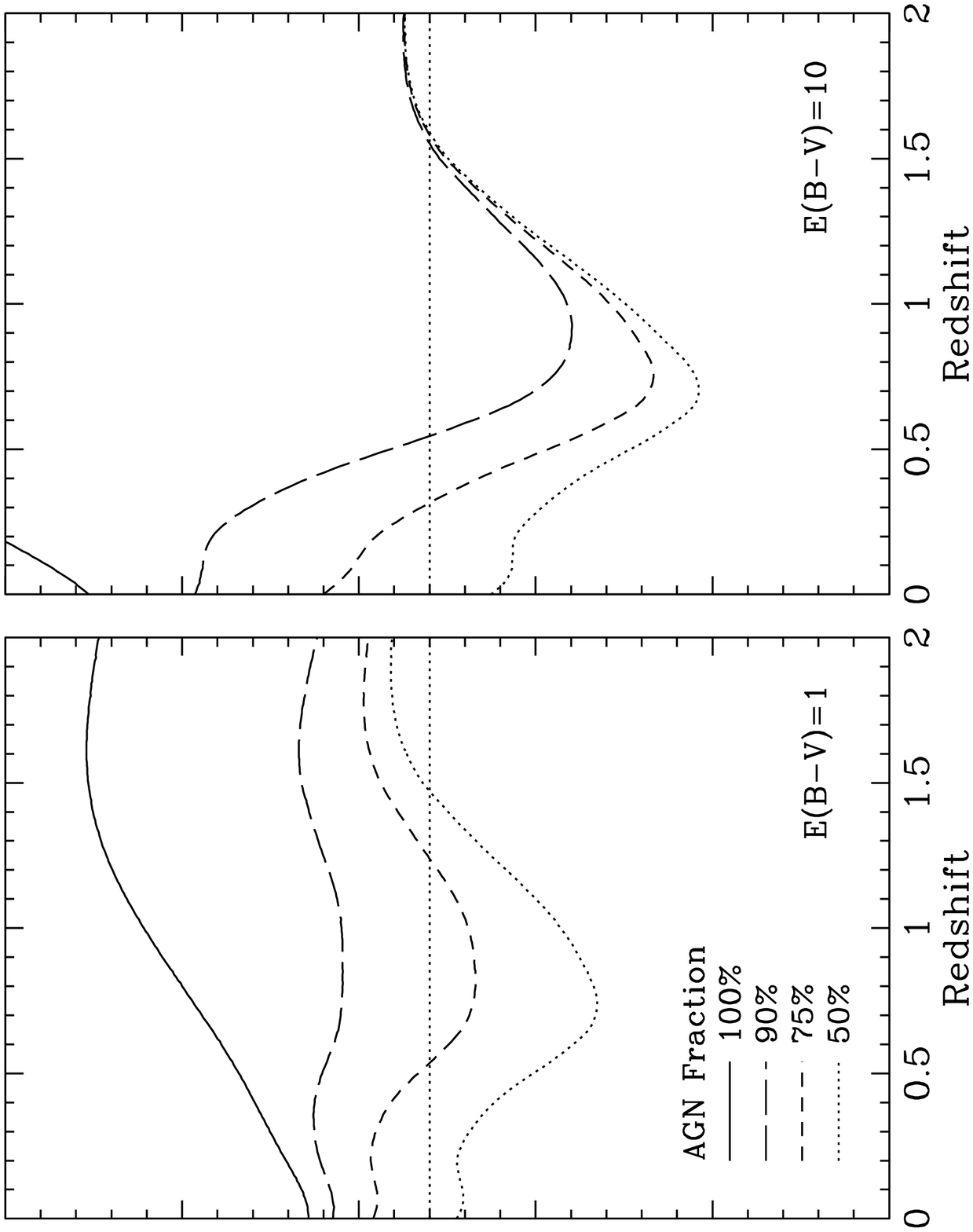}{4.5in}{-90}{60}{60}{-240}{360}
\caption{Model colors of an extincted AGN as a function of redshift
for increasing contributions of host galaxy light.  As per Fig.~1,
models use the AGN and early-type (E) galaxy templates of Assef et
al. (2010; Vega magnitudes); changing the galaxy template has minimal
effect.  The left panel shows a modestly extincted AGN with $E(B-V)
= 1$, corresponding to $N_H \sim 6 \times 10^{22}\, {\rm cm}^{-2}$,
while the right panels shows a heavily extincted AGN with $E(B-V)
= 10$, corresponding to $N_H \sim 6 \times 10^{23}\, {\rm cm}^{-2}$.
For modest levels of extinction, the results are essentially unchanged
from the right panel of Fig.~1: a simple color criterion of $W1 -
W2 \geq 0.8$ identifies AGN-dominated galaxies so long as the AGN
fraction is $\simgt 80\%$.  For heavily extincted AGN, including
Compton-thick AGN ($N_H \simgt 10^{24}\, {\rm cm}^{-2}$), the
mid-infrared emission becomes dominated by the host galaxy above
redshifts of a few tenths, essentially regardless of AGN fraction, making it difficult to distinguish such
systems from normal galaxies.
\label{fig:assef2}}
\end{figure}

Fig.~\ref{fig:assef} illustrates \wise\, selection of AGN.  As
anticipated prior to the launch of \wise\, in \citet{Ashby:09},
\citet{Assef:10} and \citet{Eckart:10}, a simple $W1 - W2$ color
cut robustly differentiates AGN from stars and galaxies.  Since the
public release of \wise\, data, several teams have also noted the
efficiency with which \wise\, identifies AGN \citep[\eg,][]{DAbrusco:12,
Edelson:12, Massaro:12}, though these analyses have used the
full four-band \wise\, photometry.  Using the empirical AGN and
galaxy spectral templates of \citet{Assef:10}, Fig.~\ref{fig:assef}
shows how $W1 - W2$ color evolves with redshift.  The left panel
considers a pure AGN template with increasing amounts of dust
extinction while the right panel considers an unobscured AGN
increasingly diluted by stellar emission (modeled with the elliptical
galaxy, or E, template).  AGN fraction refers to the fraction of
the integrated emission in the rest-frame 0.1 $-$ 30 $\mu$m range
of the unextincted templates which comes from the AGN.

Out to $z \sim 3.5$, pure AGN have red $W1 - W2$ mid-infrared colors.
Beyond this redshift, the templates become blue as the $\sim 1 \mu$m
minimum in the AGN template shifts into the $W2$ band \citep[see
also][]{Richards:06b}.  H$\alpha$ emission shifting into the $W1$
band plays an additional role in causing blue $W1 - W2$ colors for
AGN at $z \simgt 3.4$ \citep{Assef:10}.  Even modest amounts of
dust extinction redden the observed $W1 - W2$ colors for high-redshift
AGN.  Heavily extincted pure AGN are extremely red; for example, a
pure AGN reddened by $E(B-V) = 15$ has $W1 - W2 > 2$ at all
redshifts.  Mid-infrared selection of AGN is remarkably robust at
identifying pure AGN regardless of redshift.  Indeed, \citet{Blain:12}
report on the \wise\, detection of many of the highest redshift,
$z > 6$ quasars known, including the recently discovered $z = 7.085$
quasar from the United Kingdom Infrared Deep Sky Survey
\citep[UKIDSS;][]{Mortlock:11}.  That quasar, the most distant
currently known, has $W1 - W2 \sim 1.2$.

In contrast, normal galaxies and Galactic sources are unlikely to
present such red $W1 - W2$ colors.  The galaxy templates of
\citet{Assef:10} are blue in this \wise\, color combination, with
$W1 - W2 \leq 0.8$ out to $z \sim 1.2$.  Given the shallow sensitivity
of \wise, only the tip of the galaxy luminosity function will be
well-detected by \wise\, at higher redshifts, particularly when the
analysis is restricted to the very conservative, $10 \sigma$ flux
limit ($W2 \sim 15.0$) we apply in this paper.  In terms of Galactic
contamination, only the coolest brown dwarfs and the most heavily
dust-reddened stars will exhibit such red \wise\, colors.
\citet{Kirkpatrick:11} show that stars of spectral class later
(e.g., cooler) than $\sim$~T1 have $W1 - W2 \geq 0.8$; these red
colors are caused by methane absorption in the $W1$ band \citep[see
also][]{Cushing:11}.  In a high Galactic latitude survey, neither
cool brown dwarfs nor dust-reddened stars will be significant
contaminants at the flux limit of \wise.

As seen in the right panel of Fig.~\ref{fig:assef}, dilution by the
host galaxy will cause blue $W1 - W2$ colors, making less powerful
AGN no longer identifiable using this simple \wise\, color criterion.
This illustrates a powerful synergy between X-ray and mid-infrared
surveys.  While sensitive soft X-ray ($\leq 10$~keV) surveys are
quite powerful at identifying even low-luminosity AGN since stellar
processes are unlikely to power X-ray emission at luminosities
greater than $\sim 10^{42}\, {\rm erg}\, {\rm s}^{-1}$
\citep[\eg,][]{Stern:02b, Brandt:05}, such surveys are not sensitive to heavily
obscured AGN since the low energy X-rays are readily absorbed and
scattered.  This is particularly true for heavily obscured low-redshift
sources; higher redshift obscured AGN are helped by advantageous
$k$-corrections \citep[\eg,][]{Stern:02a}.  Mid-infrared surveys,
in contrast, readily identify the most heavily obscured, luminous
AGN since the obscuring material is thermally heated by the AGN and
emits relatively unimpeded by dust extinction.  However, dilution
by the host galaxy limits mid-infrared surveys from identifying low
luminosity AGN.  Optical photometric surveys are the most heavily
biased, with a sensitivity largely restricted to the least obscured,
most luminous AGN.  Using data from deep {\it Chandra} and {\it
Spitzer} imaging of targeted surveys, significant advances have
come in recent years at understanding the full census of AGN
\cite[\eg,][]{ Polletta:06, Hickox:07, Fiore:08, Gorjian:08,
Comastri:11, Hickox:11, Mullaney:11}.  Combining \wise\, with soft
X-ray data from the all-sky eROSITA telescope on {\it Spectrum
R\"ontgen Gamma (SRG)} \citep{Predehl:10}, expected to launch in
late 2013, will extend the results of these targeted surveys across
the full sky.

How will \wise\, perform at identifying the most heavily obscured,
Compton-thick AGN?  Assuming an SMC-like gas-to-dust ratio \citep[$N_H
\sim 2 \times 10^{22}\, {\rm cm}^{-2}\, {\rm mag}^{-1}$;][]{Maiolino:01},
$N_H \sim 10^{24}\, {\rm cm}^{-2}$ corresponds to $A_V \sim 50$,
or $E(B-V) \sim 15$ for $R_V \sim 3.1$ \citep[\eg,][]{Cardelli:89,
Gordon:98, York:06}.  As can be inferred from the left panel of
Fig.~\ref{fig:assef}, such heavily obscured AGN will have very red
$W1 - W2$ colors at any redshift.  However, host galaxy dilution
in these bands will become significant as extreme obscuration hides
the AGN.  As shown in Fig.~\ref{fig:assef2}, the effect is subtle
for modestly extincted sources with $N_H \simlt 6 \times 10^{22}\,
{\rm cm}^{-2}$, corresponding to $E(B-V) \simlt 1$.  Such AGN should
be readily identifiable from their $W1 - W2$ colors so long as the
AGN is bolometrically dominant.  However, the blue mid-infrared
colors of the host stellar populations across these \wise\, bands
will make low-redshift, heavily obscured AGN difficult to identify
at mid-infrared wavelengths.  At higher redshifts, $z \simgt 1.5$,
the host galaxy becomes red across these bands, but also fades below
the detection limit of \wise.   The most heavily-obscured AGN are
most likely best identified using the longer wavelength \wise\,
passbands.  However, the cost is that the diminished sensitivity
of those bands limits searches to the most luminous sources.  Indeed,
\citet{Eisenhardt:12} report on a \wise-selected source which is
undetected in $W1$ and $W2$, but has very red $W2 - W3$ colors.
Similarly selected sources are further discussed in \citet{Bridge:12}
and \citet{Wu:12}; we find only $\sim 1000$ such extreme sources
across the full sky.

This paper reports on \wise-selected AGN in the Cosmic Evolution
Survey \citep[COSMOS;][]{Scoville:07}.  We use this well studied
field, which includes deep, public, panchromatic imaging from the
radio to the X-ray in order to both establish \wise\, AGN selection
criteria and to understand the multi-wavelength properties of
\wise-selected AGN.  Our selection criterion identifies 130 AGN
candidates in COSMOS, which is sufficient for some investigations
but is too small for evolutionary studies.  A companion paper,
\citet{Assef:12}, uses the wider area Bo\"otes field in order to
investigate the luminosity distribution and evolution of \wise-selected
AGN.

This paper is organized as follows.  Section~2 discusses how the
\wise\, and COSMOS data were matched, and motivates the simple $W1
- W2 \geq 0.8$ criterion we use to identify AGN candidates.  Section~3
describes the multiwavelength properties of \wise-selected AGN,
ranging from their demographics to their {\it Hubble Space Telescope}
morphologies to their redshift distribution.  As part of this
investigation, we obtained Keck spectroscopy of mid-infrared selected
AGN candidates in the COSMOS field, described in \S~3.7.  Section~4
summarizes our results.  Since COSMOS is a well-studied field, of
interest to a broad segment of the astrophysical community, we
include an Appendix tabulating 26 additional COSMOS sources for
which we obtained spectroscopic redshifts.

Unless otherwise specified, we use Vega magnitudes throughout and
adopt the concordance cosmology, $\Omega_M = 0.3$, $\Omega_\Lambda
= 0.7$ and $H_0 = 70\, \kmsMpc$.

\section{WISE Selection of AGN in the COSMOS Field}

\subsection{Matching WISE with S-COSMOS}

The {\it Spitzer}-COSMOS survey \citep[S-COSMOS;][]{Sanders:07}
carried out a deep (620~hr), uniform survey of the full 2 deg$^2$
COSMOS field in all seven {\it Spitzer} bands (3.6, 4.5, 5.8, 8.0,
24, 70 and 160 $\mu$m).  The IRAC portion of the survey covered the
field to a depth of 1200~s in the four bluest bands of {\it Spitzer},
with $5 \sigma$ measured sensitivities ranging from $0.9 \mu$Jy at
$3.6 \mu$m to $14.6 \mu$Jy at $8.0 \mu$m.  The longer wavelength
observations, obtained with the Multiband Imaging Photometer for
{\it Spitzer} \citep[MIPS;][]{Rieke:04}, reach $5 \sigma$ sensitivities
of approximately 0.07, 8.5 and 65~mJy for the 24, 70, and 160 $\mu$m
arrays, respectively \citep{Frayer:09}.  These depths are all
considerably deeper than \wise.

We identified 6261 unique \wise\, sources with signal-to-noise ratio
$SNR \ge 10$ in \wise\, band $W2$ in a region that extends slightly
beyond the field of view (FoV) of the S-COSMOS survey.  We used a
preliminary version of the second pass data, which co-adds all
observations from the \wise\, mission.  In order to avoid spurious
and poorly photometered sources, we limited the sample to relatively
isolated sources by requiring blend flag $NB \leq 2$.  We also
avoided contaminated or confused sources by eliminating sources
whose $W1$ or $W2$ photometry was affected by diffraction spikes
($ccflag = D$), persistence ($ccflag = P$), scattered light haloes
from nearby bright sources ($ccflag = H$), or optical ghosts ($ccflag
= O$) \citep[for a detailed description of \wise\, catalog variables,
see][]{Wright:10}.  The preliminary version of the second pass data
we used double counts sources in the overlap regions between
processing stripes.  Using a 0\farcs5 match radius, we identified
duplicated sources in this preliminary catalog and only retained
the source with higher signal-to-noise ratio in $W1$.  The conservative
($10\sigma$) $W2$ depth we apply corresponds to an approximate flux
density limit of 15.05 mag ($\sim 160$ $\mu$Jy) at 4.6 $\mu$m.  Most
sources are also detected at $\geq 10\sigma$ at 3.4 $\mu$m,
corresponding to 16.45 mag ($\sim 70$ $\mu$Jy).

\begin{figure}[t!]
\plotfiddle{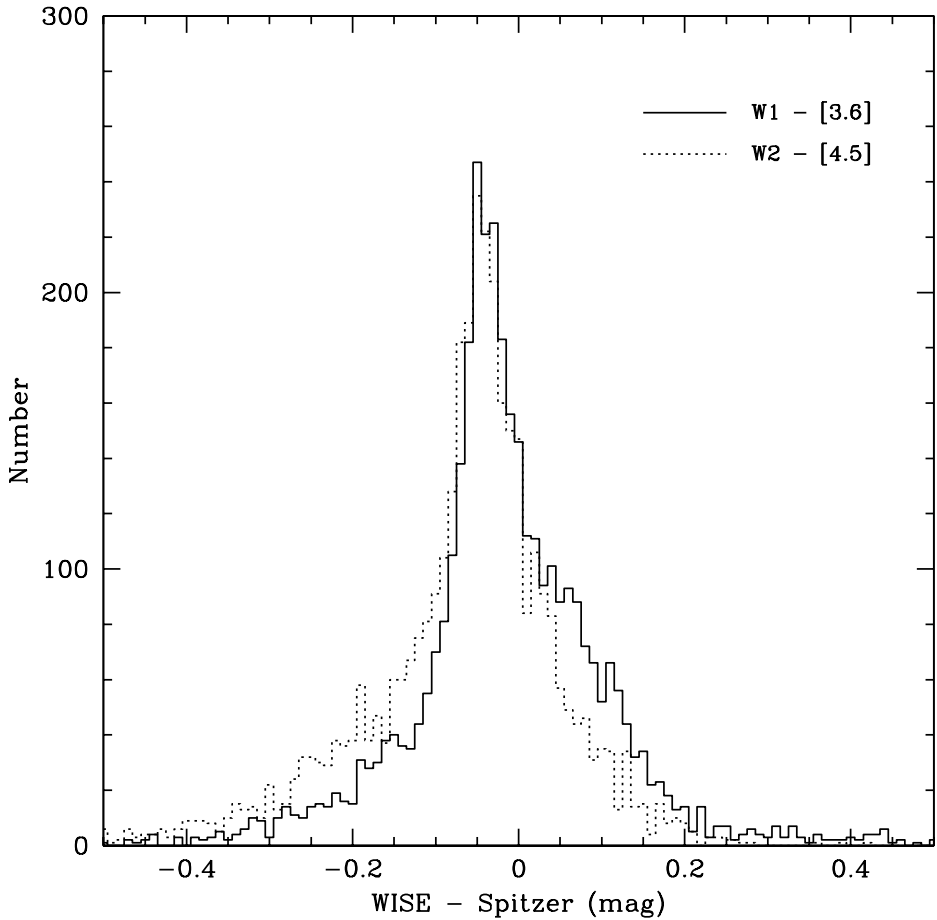}{4.5in}{0}{65}{65}{-210}{-105}
\caption{Distribution of offsets between \wise\, and S-COSMOS
photometry for the two bluest \wise\, passbands:  $W1 - [3.6]$ is
plotted as a solid line, $W2 - [4.5]$ is plotted as a dotted line.
\label{fig:magcomp}}
\end{figure}

We then identified the nearest S-COSMOS source to each \wise\,
source.  We required detections in both IRAC channel 1 (3.6
$\mu$m) and channel 2 (4.5 $\mu$m) in order to avoid the edges of
S-COSMOS which did not receive full four-band IRAC coverage.  We
also eliminated saturated stars with the requirement [3.6] $\geq
11$.  Correcting for the small mean astrometric offset between
\wise\, and S-COSMOS, $\langle \Delta {\rm R.A.} \rangle = 0\farcs108$
and $\langle \Delta {\rm Dec.} \rangle = 0\farcs008$, and requiring
a conservative 1\farcs0 matching radius, we find unique, unsaturated,
multi-band S-COSMOS identifications for 3618 \wise\, sources.  Most
of the \wise\, sources lacking S-COSMOS counterparts are from outside
the S-COSMOS FoV.  Within the area of good, unique matches, $4\%$
of \wise\, sources do not have S-COSMOS counterparts and $< 1\%$
of \wise\, sources have multiple S-COSMOS counterparts within the
1\farcs0 matching radius.  Visual inspection shows that confusion
is the source of both of these issues, with the lower resolution
\wise\, images merging multiple objects.  In the remainder of the
paper, we restrict the analysis to the $\sim 95\%$ of \wise\, sources
within the S-COSMOS area with unique, unsaturated multi-band IRAC
identifications.

Fig.~\ref{fig:magcomp} shows the measured differences between the
\wise\, and S-COSMOS photometry.  For S-COSMOS, we use aperture-corrected
2\farcs9 photometry from the public June 2007 catalog, converted
from physical units to Vega magnitudes using conversion factors
prescribed by the S-COSMOS documentation available through the
Infrared Science Archive (IRSA).  As expected given the slightly
different central wavelengths and widths of the IRAC and \wise\,
filters, we find slight median offsets between their respective
photometric measurements: $(W1 - [3.6])_{\rm med} = -0.01$ and $(W2
- [4.5])_{\rm med} = -0.07$.  The Explanatory Supplement to the
\wise\, All-Sky Data Release\footnote{See {\tt
http://wise2.ipac.caltech.edu/docs/release/allsky/expsup/index.html}.}
finds a color term from analysis of compact sources in the
\spitzer\, SWIRE {\it XMM}-LSS field, $(W1 - [3.6]) \sim 0.4 ([3.6]
- [4.5])$ (see Fig.~2 of \S VI.3.a of the Explanatory Supplement).

\begin{figure}[t!]
\plotfiddle{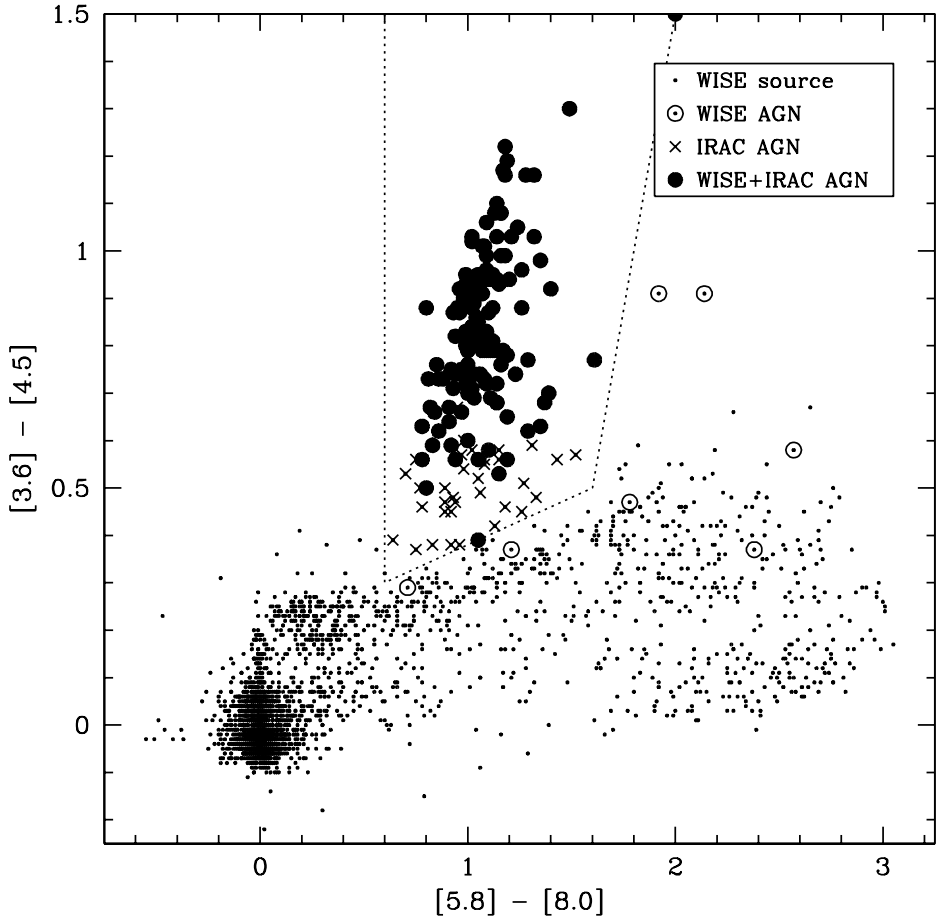}{4.5in}{0}{65}{65}{-210}{-105}
\caption{IRAC color-color diagram of \wise-selected sources in the
COSMOS field.  We only plot sources with $SNR \ge 10$ in $W1$ and
$W2$, and we require [3.6] $> 11$ to avoid saturated stars.  Sources
with $W1 - W2 \ge 0.8$ are indicated with larger circles; filled
circles indicate sources that were also identified as AGN using the
Stern et al.  (2005) mid-infrared color criteria.  Sources identified
as AGN using {\it Spitzer} criteria but not using the \wise\,
criterion are indicated with exes.
\label{fig:color_color2}}
\end{figure}

\subsection{Mid-Infrared Selection of AGN with Spitzer}

Fig.~\ref{fig:color_color2} shows the IRAC color-color diagram for
S-COSMOS sources with robust \wise\, counterparts.  The figure shows
the expected concentration of Galactic stars with Vega colors of
zero.  As discussed in \citet{Stern:07} and \citet{Eisenhardt:10},
stars warmer than spectral class T3 all have essentially Rayleigh-Jeans
continua in the IRAC passbands, leading to similar IRAC colors.
Methane absorption causes redder [3.6]$-$[4.5] colors for cooler
brown dwarfs, leading to a vertical extension above the Galactic
star locus.  Few such sources are found by \wise\, over an area as
small as COSMOS.  The sources extending to the right of the stellar
locus is dominated by low-redshift star-forming galaxies, where
polycyclic aromatic hydrocarbon (PAH) emission causes red [5.8]$-$[8.0]
colors.  Finally, as suggested in \citet{Eisenhardt:04} and discussed
in detail in \citet{Stern:05b}, the vertical extension perpendicular
to the galaxy sequence is dominated by AGN.  Indeed, \citet{Gorjian:08}
show that the majority (65\%) of X-ray sources in the XBo\"otes
survey are identified by the \citet{Stern:05b} mid-infrared criteria
\citep[see also][]{Donley:07, Eckart:10, Assef:11}.  In IRAC data
plotted to deeper depths, a highly populated second vertical sequence
is also visible to the left of the AGN sequence.  This sequence,
due to massive galaxies at $z \simgt 1.2$ \citep[e.g.,][]{Stern:05b,
Eisenhardt:08, Papovich:08}, typically outnumbers the AGN sequence
since even very shallow ($< 90$~sec) IRAC pointings easily reach
well below the characteristic brightness of early-type galaxies out
to $z \sim 2$, $m^*_{3.6} \approx 17.5$ and $m^*_{4.5} \approx 16.7$
\citep[e.g.,][]{Mancone:10}.  However, these galaxies are absent
with our conservative, $W2 \simlt 15$ magnitude cut in the much
shallower \wise\, data.

In the following analysis we will adopt the \citet{Stern:05b}
mid-infrared AGN sample as the `truth sample' in order to explore
potential \wise\, AGN selection criteria.  The \citet{Stern:05b}
method for selecting AGN was one of the first methods devised to
identify AGN using \spitzer\, data and has been extensively used
by other workers in the field.  However, like all AGN selection
criteria, it is not without some shortcomings, highlighted below.

{\it X-ray selected AGN missed by mid-infrared selection:}  As
pointed out by numerous authors \citep[\eg,][]{Barmby:06, Cardamone:08,
Brusa:09}, many X-ray sources have mid-infrared colors consistent
with normal galaxies, and thus are missed by the mid-infrared
AGN color criteria.  As first pointed out by \citet{Donley:07} and
further expanded upon by \citet{Eckart:10}, the fraction of X-ray
sources identified as mid-infrared AGN increases strongly with X-ray
luminosity.  For example, \citet{Donley:07}, using data from the
Ms \chandra\, Deep Field-North \citep{Alexander:03}, find that the
mid-infrared selection efficiency increases from $\sim 14\%$ at
$L_{\rm 0.5-8\, keV} < 10^{42}\, {\rm erg}\, {\rm s}^{-1}$ to $100
\%$ at $L_{\rm 0.5-8\, keV} > 10^{44}\, {\rm erg}\, {\rm s}^{-1}$.
So while low-luminosity AGN are, unsurprisingly, missed by the
mid-infrared color selection criteria  (\eg, Figs.~\ref{fig:assef} and
\ref{fig:assef2}), such criteria appear remarkably robust at
identifying the most luminous AGN in the universe.

{\it Contamination by star-forming galaxies:}  Several authors have
also pointed out that the mid-infrared color cuts proposed by
\citet{Lacy:04} and \citet{Stern:05b} extend into regions of color
space populated by star-forming galaxies \citep[\eg,][]{Barmby:06,
Donley:08, Park:10}.  In order to minimize such contamination,
mid-infrared power-law selection has been suggested
\citep[\eg,][]{AlonsoHerrero:06}, though \citet{Donley:12} notes
that systematic photometric errors from IRAC are often underestimated
\citep[\eg,][]{Reach:05}, making power-law selection more vulnerable
to the quality of the mid-infrared photometry than simple color-color
cuts.  \citet{Donley:12} investigates contamination by star-forming
galaxies using a combination of galaxy templates and real data from
pure starbursts identified from \spitzer\, IRS spectroscopy.  A
strong conclusion from this work is that mid-infrared color selection,
particularly using the \citet{Stern:05b} criteria, has minimal
contamination from purely star-forming galaxies below a redshift
of $z \sim 1$.

{\it Contamination by high-redshift galaxies:}  The \citet{Lacy:04}
and \citet{Stern:05b} criteria for identifying AGN based on their
mid-infrared colors were empirically derived from shallow, wide-area
\spitzer\, data.  In this limit, the criteria work extremely well.
In deeper mid-infrared data, however, significant contamination from
faint, high-redshift galaxies becomes problematic and we recommend
use of the revised IRAC selection criteria of \citet{Donley:12} for
deep IRAC data.  

{\it Completeness vs. reliability:}  Finally, in choosing between
the IRAC color criteria of \citet{Lacy:04} and \citet{Stern:05b},
we have opted for the latter.  As pointed out by numerous authors
\citep[\eg,][]{Eckart:10, Donley:12}, the less selective \citet{Lacy:04}
criteria have higher completeness at the cost of reliability:  many
more normal galaxies are identified using those criteria, yielding
more significant contamination.  With the primary goal of identifying
a clean sample of powerful AGN, we therefore adopt the higher
reliability IRAC color criteria of \citet{Stern:05b}.

In summary, we adopt the \citet{Stern:05b} mid-infrared-selected
AGN candidates as the `truth sample' for analyzing \wise\, selection
of AGN.  Foremost, identifying a robust truth sample identified at
similar wavelength makes logical sense.  Radio selection would miss
$\sim 90\%$ of AGN, while optical and X-ray selection would miss
the heavily obscured AGN which are identified by mid-infrared
selection but largely missed by the current generation of optical and X-ray surveys.  In
principle, a hybrid selection could be adopted, such as identifying
all AGN candidates with $L_{\rm AGN}$ greater than some value.
However, a `truth sample' identified in that manner would be
vulnerable to spectroscopic incompleteness.  

Despite these caveats to mid-infrared selection, we show that the
\citet{Stern:05b} criteria are quite robust at the shallow mid-infrared
depths of \wise.  This method identifies the most luminous
X-ray-selected AGN at high completeness.  It also identifies a much
higher surface density of AGN than optical and X-ray surveys of
comparable depth due to an increased sensitivity to obscured AGN.
Illustrating this point, \citet{Hickox:07} and \citet{Eckart:10}
show that mid-infrared AGN candidates individually undetected at
high energy are well detected in stacking analyses and reveal a
harder-than-average X-ray spectrum, implying significant obscuration.
Optical spectroscopy of IRAC-selected AGN candidates also typically
reveal type~2 AGN spectra (Stern \etal, in prep.).  Finally, in
shallow mid-infrared data, particularly at the depth of \wise, the
\citet{Stern:05b} criteria suffer from minimal contamination by
Galactic stars, starburst galaxies, or high-redshift galaxies.

\subsection{Mid-Infrared Selection of AGN with WISE}

\begin{figure}[t!]
\plotfiddle{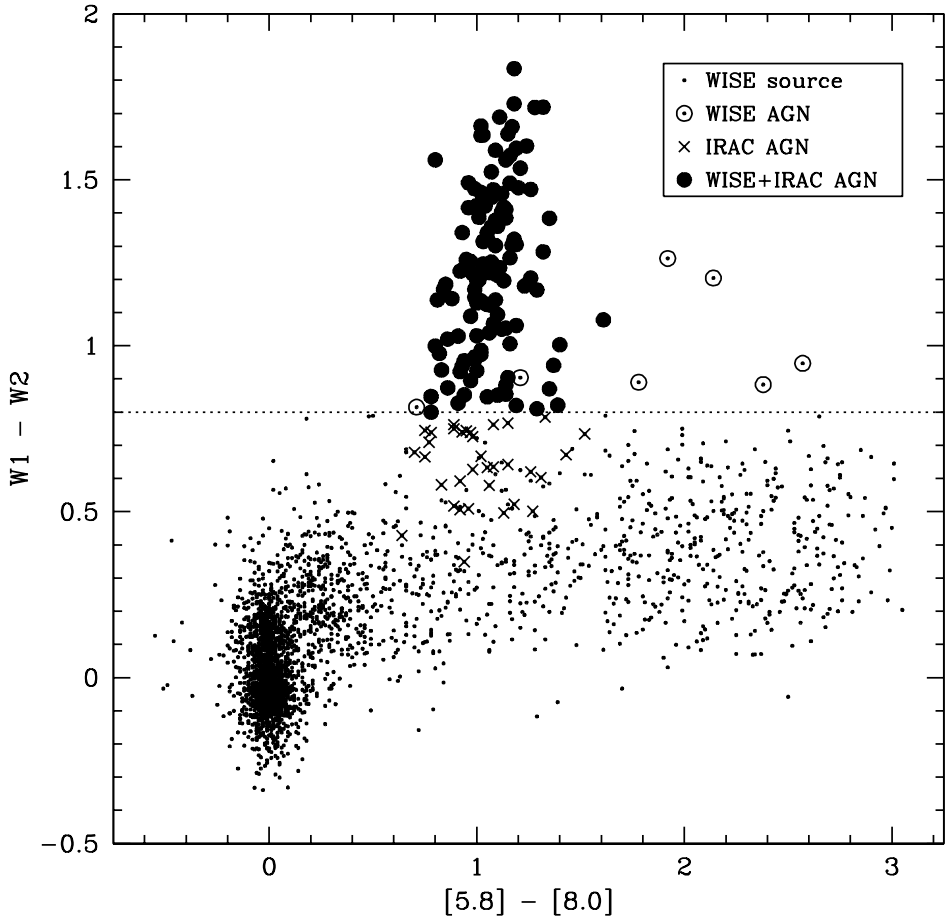}{4.5in}{0}{65}{65}{-210}{-105}
\caption{Mid-infrared color-color diagram of \wise-selected sources
in the COSMOS field, with \wise\, $W1-W2$ plotted against IRAC
$[5.8]-[8.0]$.  Symbols are as in the previous figure.
\label{fig:color_color1}}
\end{figure}

Fig.~\ref{fig:color_color1} shows a hybrid mid-infrared color-color
diagram.  Rather than plotting IRAC [3.6]$-$[4.5] color along the
vertical axis, we plot $W1 - W2$.  We see similar trends to the
IRAC-only diagram (Fig.~\ref{fig:color_color2}), with a stellar
locus at zero color, a horizontal sequence of low-redshift galaxies,
and a vertical AGN sequence perpendicular to the galactic sequence.
While mid-infrared selection of AGN in even very shallow {\it
Spitzer} pointings required the longer wavelength IRAC passbands
to differentiate AGN from high-redshift ($z \simgt 1.3$) massive
galaxies, \wise\, is able to robustly identify AGN with just $W1$
and $W2$ \citep[e.g., see][]{Ashby:09, Assef:10, Eckart:10}.  Note
that these are the two most sensitive \wise\, passbands, with the
highest source counts and the best spatial resolution.

\begin{figure}[t!]
\plotfiddle{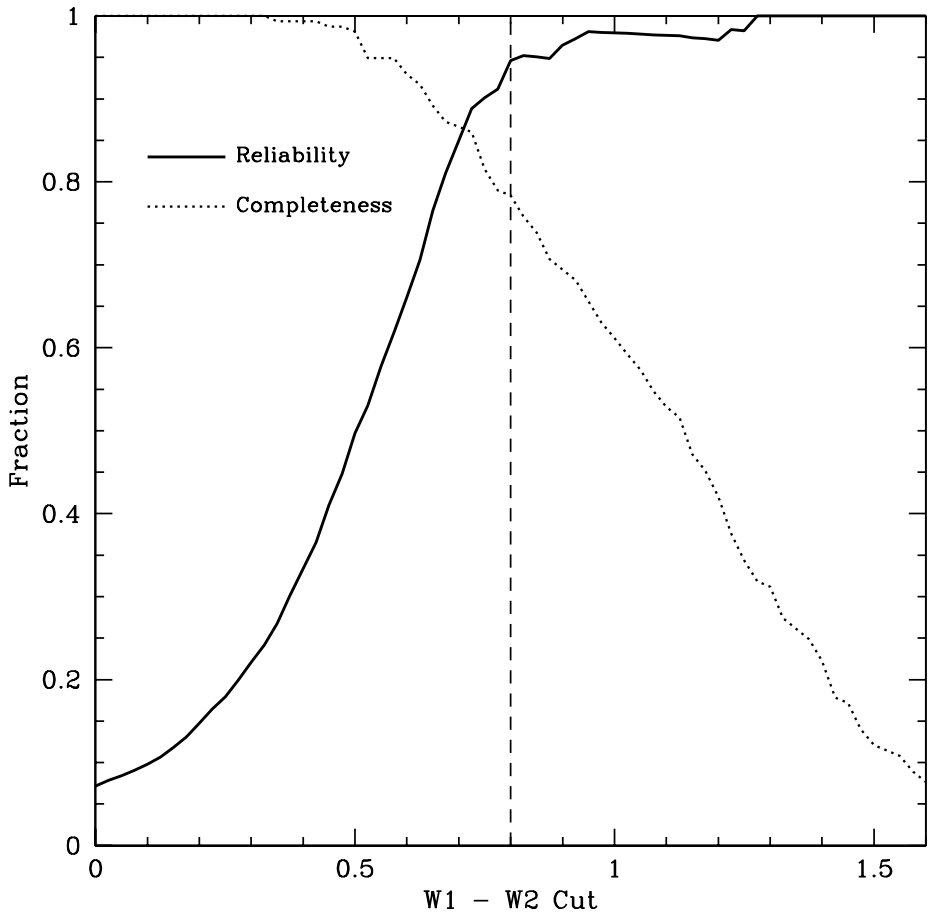}{4.5in}{0}{65}{65}{-210}{-105}
\caption{Reliability (solid line) and completeness (dotted line) of \wise\,
AGN selection as a function of a simple $W1 - W2$ color selection.
Blue cuts have very high completeness --- e.g., select all sources
which are identified as mid-infrared AGN candidates according to
the Stern et al. (2005) IRAC criteria.  However, blue cuts have
poor reliability, selecting many sources whose mid-infrared colors
suggest they are normal galaxies.  Likewise, red cuts robustly
select AGN with few contaminants, but have low completeness.  Using
the criteria $W1 - W2 \ge 0.8$ offers both high completeness (78\%)
and high reliability (95\%).
\label{fig:completeness_reliability}}
\end{figure}

We have explored how robustly \wise\, identifies AGN using a simple
$W1 - W2$ color cut.  Of the 3618 sources in our cross-matched
\wise-S-COSMOS catalog, 157 are mid-infrared AGN according to the
\citet{Stern:05b} criteria.  We consider this the truth sample and
Fig.~\ref{fig:completeness_reliability} shows the completeness and
reliability of \wise\, AGN selection as a function of $W1 - W2$
color cut.  Prior to the launch of \wise, \citet{Ashby:09} suggested
$W1 - W2 \ge 0.5$ would robustly identify AGN while \citet{Assef:10}
suggested a color cut of $W1 - W2 \ge 0.85$.  Considering the former,
while this criterion is highly (98\%) complete at identifying the
AGN sample, it suffers from significant contamination from non-active
sources (see Fig.~\ref{fig:completeness_reliability}; only 50\% of
sources appear active according to the IRAC criteria).  This is
likely, in part, due to the significantly better performance of
\wise\, compared to the prelaunch predictions:  \citet{Mainzer:05}
reported 5$\sigma$ point source sensitivity requirements of 120
$\mu$Jy at 3.4 $\mu$m and 160 $\mu$Jy at 4.6 $\mu$m, while we are
finding 10$\sigma$ point source sensitivities of 70 $\mu$Jy at 3.4
$\mu$m and 160 $\mu$Jy at 4.6 $\mu$m.  Analysis of
Fig.~\ref{fig:completeness_reliability} suggests a color cut at $W1
- W2 = 0.8$ offers an extremely robust AGN sample which is still
highly complete.  For some uses, a slightly less conservative color
cut at $W1 - W2 = 0.7$ might be preferable, providing a powerful
compromise between completeness and reliability for \wise\, AGN
selection.

Using $W1 - W2 \ge 0.8$ to select AGN candidates, we identify 130
potential candidates, of which 123 are AGN according to their IRAC
colors (95\% reliability, 78\% completeness).  The less conservative
color cut at $W1 - W2 \ge 0.7$ identifies 160 candidates, of which
136 are AGN according to their IRAC colors (85\% reliability, 87\%
completeness) --- \eg, this less restrictive color cut identifies
$\sim 10 \%$ more AGN at the cost of tripling the number of
contaminants.  As seen in Fig.~\ref{fig:color_color2}, several of the
``mis-identified'' AGN candidates have colors very close to the
\citet{Stern:05b} criteria.  In fact, as discussed in \S~3.7, four of the seven
``contaminants'' for the $W1 - W2 \ge 0.8$ AGN selection have spectroscopic
redshifts, two of which are broad-lined quasars.  This implies that
the above reliability numbers are likely conservative, though complete
spectroscopy will be required to determine what fraction of the
IRAC-selected AGN are, in fact, not active. 
As seen in Fig.~\ref{fig:color_color1},
mis-identified AGN candidates often have much redder $[5.8]-[8.0]$ colors,
suggestive of low-redshift galaxies with PAH emission.  Potential
Galactic contaminants are brown dwarfs (cooler than spectral class
T3) and asymptotic giant branch (AGB) stars, both of which have
much lower surface densities than AGN at the depths probed and are
not expected to be a significant contaminant, particularly in
extragalactic pointings.


In the following section we use the extensive publicly available
data in the COSMOS field to explore the demographics, multiwavelength
properties and redshift distribution of \wise-selected AGN using
the simple $W1 - W2 \ge 0.8$ color criterion.

\section{Properties of WISE AGN}

\subsection{Demographics of WISE AGN}

The effective area of the S-COSMOS survey is 2.3 deg$^2$ per passband
after removal of poor quality regions around saturated stars and
the field boundary \citep{Sanders:07}.  Since approximately 8\% of
the field is only covered by two of the four IRAC passbands, this
implies a four-band effective area of approximately 2.1 deg$^2$.
Our simple $W1 - W2$ color criterion identified 130 AGN candidates
in this area, implying a surface density of $61.9 \pm 5.4$
\wise-selected AGN candidates per deg$^2$, 5\% of which are expected
to be contaminants.

For comparison, the Sloan Digital Sky Survey (SDSS) quasar selection
algorithm \citep{Richards:02b} targeted ultraviolet excess quasars
to $i^* = 19.1$ (AB mag; 13.0 targets per deg$^2$) and higher
redshift ($z \simgt 3$) quasars to $i^* = 20.2$ (AB mag; 7.7 targets
per deg$^2$), yielding a combined list of 18.7 candidates per
deg$^2$.  These depths are comparable to \wise\, depths for type~1
quasar selection \citep[\eg,][]{Assef:10}.  The SDSS algorithm is
expected to provide over 90\% completeness from simulated type 1
quasar spectra, although at the cost of lower reliability.  The
overall efficiency (quasars / quasar candidates) was only 66\% from
initial test data over 100 deg$^2$, with the contaminants evenly
split between galaxies and Galactic stars.  Importantly, optical/UV
quasar selection methods are hampered at certain redshifts, especially
high ones, where the stellar locus overlaps the quasar locus in
color-color space.  More sophisticated approaches, such as that
presented in \citet{Bovy:11}, work better than the old two-color
or three-color cuts, though they still have issues.  Other methods,
such as variability \citep[\eg,][]{PalanqueDelabrouille:11}, also
do better, but require more elaborate input data sets.  \wise\,
selection is less affected by these problems, and has a much flatter
selection function as a function of redshift than traditional
color-selected UV-excess methods.  In particular, our simple $W1 -
W2$ selection is expected to have 78\% completeness and 95\%
reliability assuming that the \citet{Stern:05b} mid-infrared selection of AGN candidates
from the deeper S-COSMOS data is 100\% reliable.

\subsection{Mid-Infrared Properties of WISE AGN}

We have studied the longer wavelength properties of \wise-selected
AGN with an eye towards investigating whether the inclusion of $W3$
or $W4$ would allow for a more robust \wise\ selection of AGN, such
as the wedge in $W1-W2$ vs. $W2-W3$ color-color space presented in
\citet{Jarrett:11}.  Our simple $W1 - W2$ color criterion identified
130 AGN candidates in COSMOS, of which 24 (18\%) are detected in
$W3$ ($\geq 10\sigma$) and only 3 (2\%) are detected in $W4$ ($\geq
10\sigma$).  If we instead use a less conservative $5\sigma$ detection
threshold, we find that 78 (60\%) are detected in $W3$ and 17 (13\%)
are detected in $W4$.  These percentages are essentially unchanged
when we consider the 123 robust AGN candidates identified by both
the \wise\ and IRAC selection criteria: 24 (20\%) are detected in
$W3$ ($\geq 10\sigma$) and 3 (2\%) are detected in $W4$ ($\geq
10\sigma$).  Using the $5\sigma$ detection threshold, these numbers
increase to 74 (60\%) being detected in $W3$ and 17 (14\%) being
detected detected in $W4$.

This implies that including the longer wavelength \wise\ data
increases the reliability of the AGN selection, but at the cost of
a significant decrease in completeness.  For example, requiring a
$10\sigma$ detection in $W3$ in addition to the $W1 - W2$ color
criterion provides a surface density of only 11 \wise-selected AGN
candidates per deg$^2$, but the reliability is expected to be $\sim
100\%$.  Using the less conservative requirement of a $5\sigma$
detection in $W3$ provides a surface density of 37 \wise-selected
AGN candidates per deg$^2$ with a reliability of $95\%$ -- \eg, the
same reliability as our original $W1 - W2 \geq 0.8$ color cut.  Any
of these mid-infrared selection criteria compare quite favorably
with the SDSS quasar selection in terms of surface density,
completeness and reliability.  However, in what follows we rely on
selecting AGN using only the single $W1 - W2 \geq 0.8$ color
criterion, as this selection, relying on the most sensitive \wise\
passbands, identifies a much larger candidate AGN population, 2 to
5 times larger than the samples that also require $W3$ detections.

Finally, we note that the depth of the \wise\, survey varies strongly
with ecliptic latitude.  The COSMOS field was selected to be at low
ecliptic latitude in order to make it accessible from both hemispheres,
and thus is close to the minimum depth of the \wise\, survey with
a coverage of 12 frames (11 sec each).  Higher latitude fields have
deeper data, reaching a coverage of 250 frames at the ecliptic poles
\citep{Jarrett:11}.  This increases the surface density of AGN
candidates, but with a decrease in robustness since deeper pointings
will detect massive galaxies at $z \simgt 1$ which have similar $W1
- W2$ colors to AGN.  The inclusion of the deeper $W3$ data in such
fields could be used to separate normal galaxies from AGN.  \wise\,
selection of AGN in deeper, higher latitude fields is addressed
more thoroughly in \citet{Assef:12}.

\subsection{X-Ray Properties of WISE AGN}

The COSMOS field has been observed by both the {\it Chandra X-Ray
Observatory} \citep{Elvis:09} and {\it XMM-Newton} \citep{Hasinger:07}.
In the following two subsections, we examine the X-ray properties
of \wise-selected AGN using each of these surveys.

\begin{figure}[t!]
\plotfiddle{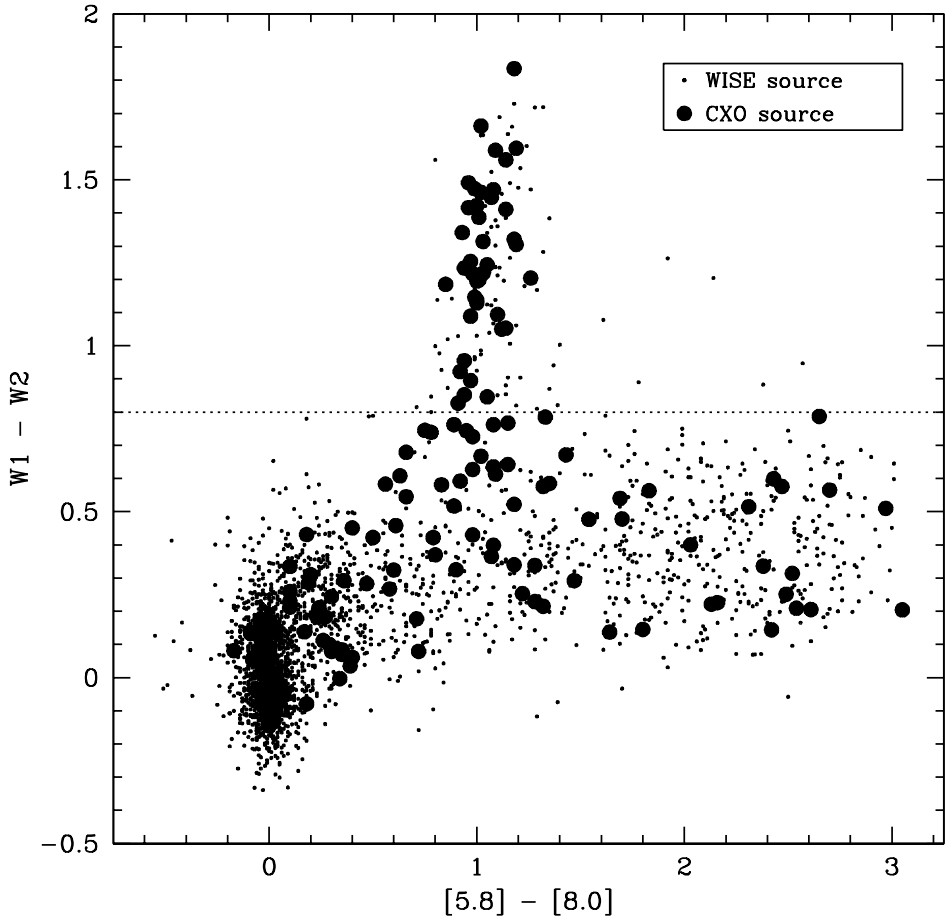}{4.5in}{0}{65}{65}{-210}{-105}
\caption{Mid-infrared color-color diagram of \wise-selected sources
in the COSMOS field, with \wise\, $W1-W2$ plotted against IRAC
$[5.8]-[8.0]$.  Many of the WISE sources lacking {\it Chandra}
counterparts are from outside the field-of-view of the C-COSMOS
survey.
\label{fig:color_colorCXO}}
\end{figure}

\subsubsection{Chandra Observations}

The \chandra-COSMOS Survey \citep[C-COSMOS;][]{Elvis:09} is a large,
1.8~Ms, \chandra\, program that imaged the central 0.5~deg$^2$ of
the COSMOS field with an effective exposure time of 160~ks per
position and the outer 0.4~deg$^2$ with an effective exposure time
of 80~ks per position.  The corresponding point source depths in
the deeper portion of the survey are $1.9 \times 10^{-16}\, \ergcm2s$
in the soft ($0.5 - 2$~keV) band, $7.3 \times 10^{-16}\, \ergcm2s$
in the hard ($2 - 10$~keV) band and $5.7 \times 10^{-16}\, \ergcm2s$
in the full ($0.5 - 10$~keV) band, where these depths assume an
average X-ray power law index $\Gamma = 1.4$.  C-COSMOS detected
1761 reliable X-ray point sources (catalog ver.2.1; spurious
probability $< 2 \times 10^{-5}$).  We use a 2\farcs5 matching
radius to cross-identify the \wise\, and \chandra\, sources.

A total of 167 of the \chandra\, X-ray sources have \wise\,
counterparts (Fig.~\ref{fig:color_colorCXO}).  Most have relatively
blue $W1 - W2$ colors and are likely associated with low-redshift
galaxies harboring low-luminosity AGNs; such sources are common in
deep X-ray observations \citep[e.g.,][]{Brandt:05}.  A concentration
of X-ray sources near zero mid-infrared colors are predominantly
low-redshift ($z \simlt 0.6$) early-type galaxies:  such galaxies
have little or no star formation and therefore lack the dust and
PAH emission which causes a horizontal extension in this mid-infrared
color-color space.  Finally, deep \chandra\ surveys are also sensitive
to X-ray emission from low-mass Galactic stars, which likewise
reside in this same region of color-color space \citep[\eg,][]{Stern:02b}.

More interesting is the vertical extension seen in
Fig.~\ref{fig:color_colorCXO}:  41 of the \wise-selected AGN
candidates have C-COSMOS counterparts.  Most of the \wise$+$IRAC
AGN candidates lacking \chandra\, counterparts in
Fig.~\ref{fig:color_colorCXO} are from outside the C-COSMOS FoV.
Six (\eg, 13\%) of the \wise-selected AGN candidates whose IRAC
colors are consistent with the \citet{Stern:05b} AGN selection
criteria were observed by, but {\em not} detected by \chandra.  We
list these sources in Table~\ref{table:chandra}.  All are at least
9\arcsec\, from the nearest \chandra\, source.  These mid-infrared
sources are strong candidates for Compton-thick AGN (\eg, $N_H \geq
10^{24}\, {\rm cm}^{-2}$):  sources with so much internal absorption
that their X-ray emission below 10~keV is heavily absorbed, undetected
in the $\sim 100$~ks {\it Chandra} observations.  The absorbing
material, however, is heated up and is easily detected in $\sim
100$~s integrations with \wise.  We also note, as expected, that
the small number of \wise\, AGN contaminants with $W1 - W2 \geq
0.8$ but whose IRAC colors are {\em not} consistent with an AGN are
undetected by \chandra.

\begin{figure}[t!]
\plotfiddle{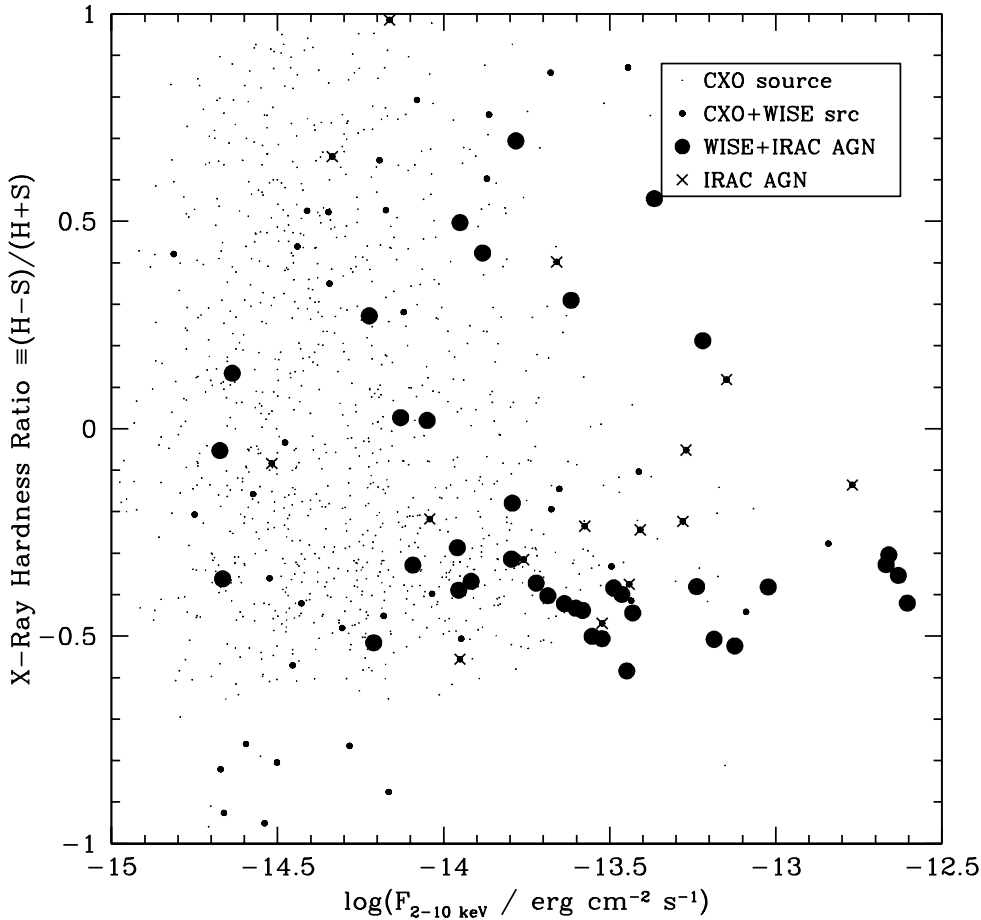}{4.5in}{0}{65}{65}{-210}{-105}
\caption{X-ray hardness ratio $HR \equiv (H-S)/(H+S)$, where $H$
($S$) is the number of detected hard (soft) X-ray counts, plotted
against hard X-ray flux.  All X-ray data is from \chandra; at the 
depth of these data, most X-ray sources are expected to be AGN.  We
identify no \chandra\, counterparts for the \wise-selected AGN
candidates whose IRAC colors are indicative of being normal,
non-active galaxies.  Note that many (but not all) of the brightest
hard X-ray sources are identified as AGN candidates by both \wise\,
and IRAC.
\label{fig:chandraflux}}
\end{figure}

Fig.~\ref{fig:chandraflux} shows the relationship between hard X-ray
($2 - 10$~keV) fluxes of S-COSMOS sources and their hardness ratios
$HR \equiv (H - S) / (H + S)$, where $H$ and $S$ are the numbers
of hard and soft X-ray photons detected, respectively.  We compare
the full sample (smallest black dots) to the subset with \wise\,
counterparts (small black circles) to the smaller set of sources
with IRAC and/or \wise\, colors indicative of AGN activity (larger
symbols).  Note that many of the brightest X-ray sources in the
field are identified as AGN candidates by \wise, regardless of their
hardness ratio.


\scriptsize
\begin{deluxetable}{cccccccl}
\tablecaption{\wise$+$IRAC AGN candidates undetected by \chandra.}
\tablehead{
\colhead{\wise\, ID} &
\colhead{$i$} &
\colhead{$W1$} &
\colhead{$W2$} &
\colhead{$W1 - W2$} &
\colhead{[5.8]$-$[8.0]} &
\colhead{$z$} &
\colhead{Notes}}
\startdata
J095855.40+022037.4 & 19.43 & 15.41 & 14.59 & 0.82 & 1.39 & [0.38] & bright galaxy \\
J095937.35+021905.9 & 22.30 & 16.00 & 14.70 & 1.30 & 1.17 & 0.927  & \\
J100006.19+015535.3 & 20.90 & 15.53 & 14.19 & 1.34 & 1.05 & 0.661  & \\
J100043.70+014202.5 & 21.77 & 15.89 & 14.89 & 1.00 & 1.40 & 0.741  & \\
J100046.91+020726.5 & 21.86 & 14.87 & 13.14 & 1.73 & 1.18 & 1.158  & faint \\
J100135.61+022104.8 & 25.11 & 16.98 & 14.97 & 2.01 & 1.49 &\nodata & faint \\
%
\enddata
\label{table:chandra}
\tablecomments{Bracketed redshift indicates photometric redshift.}
\end{deluxetable}
\normalsize

\begin{figure}[t!]
\plotfiddle{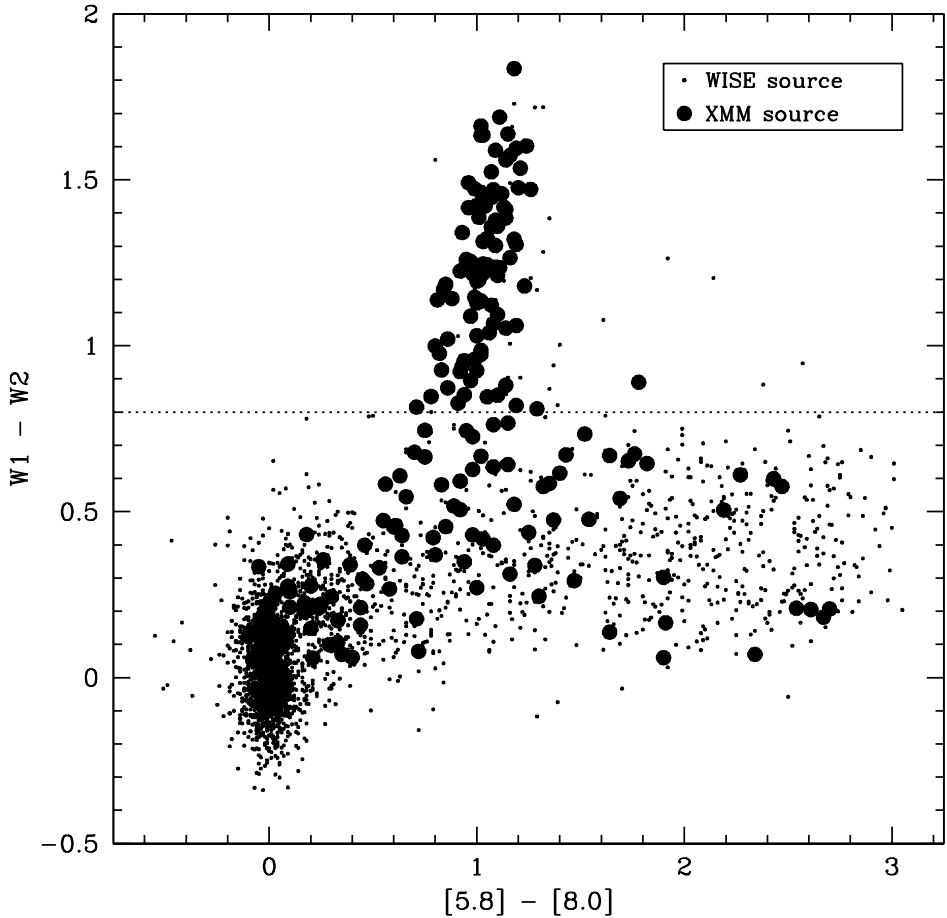}{4.5in}{0}{65}{65}{-210}{-105}
\caption{Mid-infrared color-color diagram of \wise-selected sources
in the COSMOS field, with \wise\, $W1-W2$ plotted against IRAC
$[5.8]-[8.0]$.  Dots show all \wise\, sources in the field, larger
filled circles show \wise\, sources with {\it XMM} counterparts.
\label{fig:color_colorXMM}}
\end{figure}

\begin{figure}[t!]
\plotfiddle{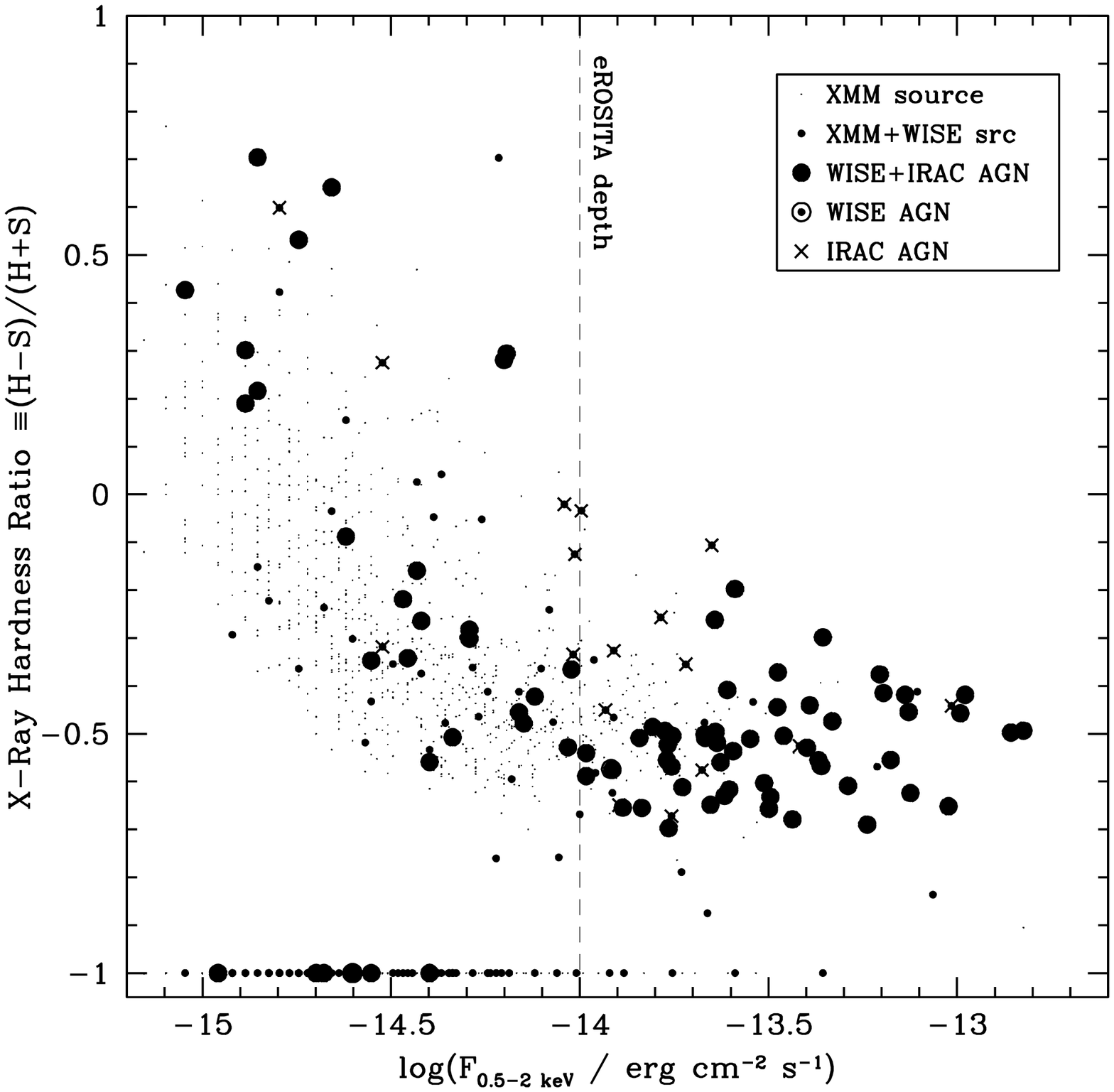}{2.5in}{0}{42}{42}{-260}{-95} 
\plotfiddle{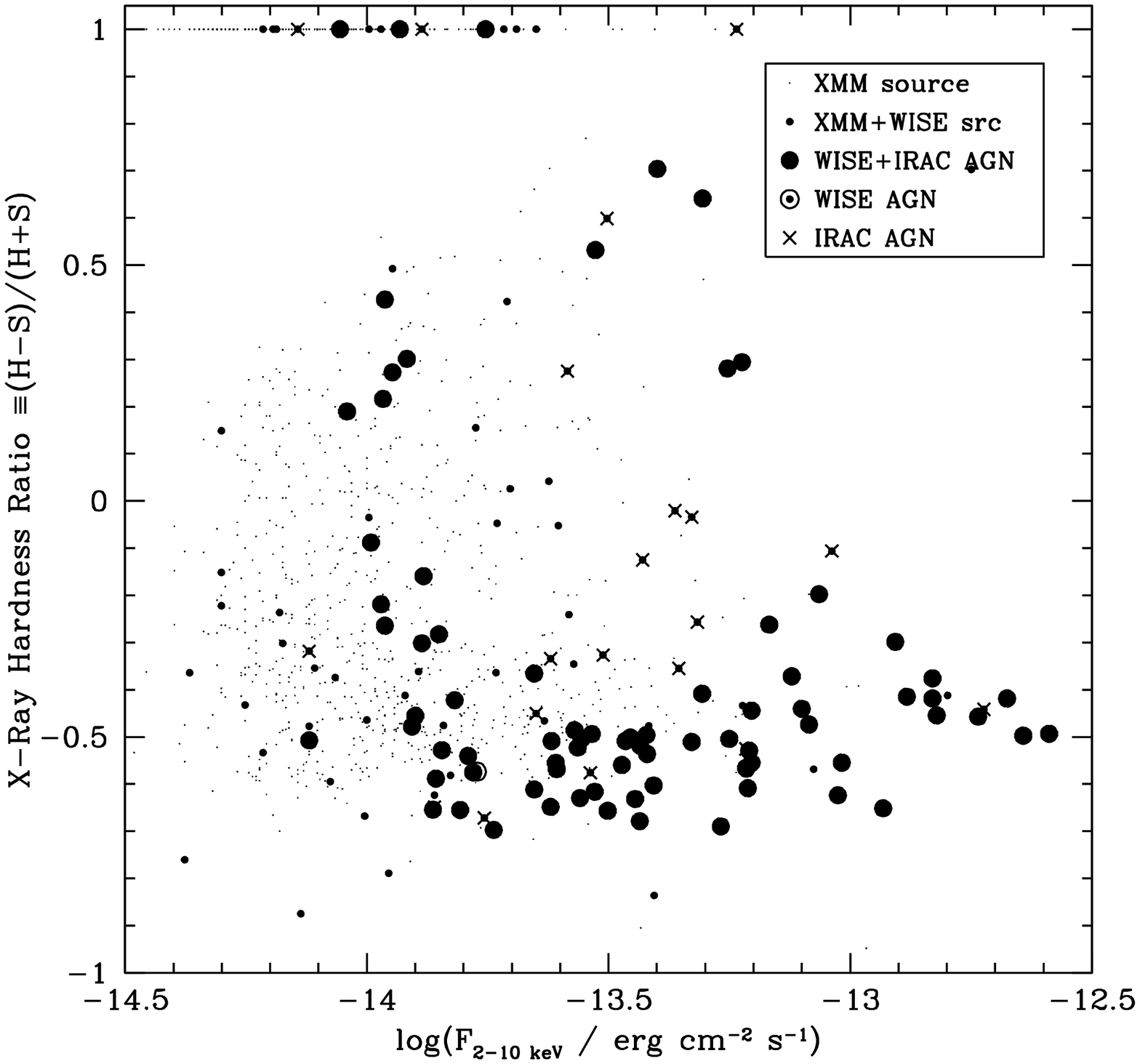}{0.0in}{0}{42}{42}{00}{-65} 
\caption{X-ray hardness ratio $HR$ from \xmm\, plotted against X-ray
flux (right: soft-band, $0.5 - 2$~keV; left:  hard-band, $2 -
10$~keV).  Note that many (but not all) of the brightest hard X-ray
sources are identified as AGN candidates by both the \wise\, criterion
and by their IRAC colors (large black circles).  The vertical dashed
line in the left panel shows the expected soft-band point source
sensitivity of the all-sky eROSITA survey; the hard-band point
source sensitivity corresponds approximately to the right-hand axis
of the high-energy panel.  While the brightest soft X-ray sources
are expected to also be identified by eROSITA, many \wise-selected
AGN candidates are below the eROSITA flux limits.  Conversely, eROSITA is
expected to identify many lower luminosity AGN that are not identified
as AGN candidates by \wise.
\label{fig:xmmflux}}
\end{figure}

\subsubsection{XMM-Newton Observations}

The \xmm\, wide-field survey of the COSMOS field \citep[{\it
XMM}-COSMOS;][]{Hasinger:07, Brusa:10} observed the entire 2 deg$^2$
COSMOS field to medium depth ($\sim 60$~ks).  The survey detected
nearly 2000 X-ray sources down to limiting fluxes of $\sim 5 \times
10^{-16} \ergcm2s$ in the $0.5 - 2$~keV (soft) band and $\sim 3
\times 10^{-15} \ergcm2s$ in the $2 - 10$~keV (hard) band.  Thus,
{\it XMM}-COSMOS covers a wider area than C-COSMOS, albeit to
shallower depth.  We use the November 2008 \xmm\, point-like source
catalog, available thru IRSA, which contains 1887 sources.  Using
a 3\farcs5 matching radius --- slightly larger than used for C-COSMOS
to account for the poorer spatial resolution of \xmm\, --- we
identify \wise\, counterparts for 244 {\it XMM}-COSMOS sources, of
which 92 have $W1 - W2 \geq 0.8$.  The mid-infrared color distribution
of XMM sources is similar to what was seen for C-COSMOS, though
more of the mid-infrared AGN candidates have X-ray detections
courtesy of the wider spatial coverage of this survey
(Fig.~\ref{fig:color_colorXMM}).  Thirty-three \wise\, sources whose IRAC
colors suggest an active nucleus (out of 123) are undetected by
\xmm, though seven are from outside the \xmm\, coverage; the other
26 are listed in Table~\ref{table:xmm}.  All six of the sources
from Table~\ref{table:chandra} remain undetected by \xmm.


Therefore, 75\% of the \wise-selected AGN candidates have \xmm\,
counterparts, but that still leaves a significant number of
\wise-selected AGN --- including those whose IRAC colors indicate
an AGN --- that are {\it undetected} by \xmm.  We also note that
two of the sources with $W1 - W2 \geq 0.8$ but outside of the
\citet{Stern:05b} IRAC wedge are detected in the X-rays; both have
IRAC colors very close to the wedge defined in that paper.

Similar to the \chandra\, results, we find that the \wise-selected
AGN are brighter and have softer spectra than typical \xmm\, sources
in {\it XMM}-COSMOS.  However, Fig.~\ref{fig:xmmflux} also shows
quite clearly that the brightest X-ray sources tend to be identified
as AGN candidates from their \wise\, colors, regardless of X-ray
hardness ratio.  For comparison, we also plot the expected point
source sensitivity of the all-sky eROSITA telescope.  The brightest
soft X-ray sources, $F_{\rm 0.5-2 keV} \simgt 2 \times 10^{-14}\,
{\rm erg}\, {\rm cm}^{-2}\, {\rm s}^{-1}$, will basically all already
have been identified as AGN candidates by \wise.  At the sensitivity
limit of eROSITA, however, large numbers of X-ray sources are
expected that are not identified by \wise; these are likely lower
luminosity AGN at lower redshifts \citep[\eg,][]{Eckart:10, Donley:12}.
\wise\, also detects a significant population of fainter, harder
spectrum X-ray sources, below the sensitivity limit of eROSITA.
These results, particularly Fig.~\ref{fig:xmmflux}, emphasizes the
complementarity of X-ray and mid-infrared AGN selection:  each
selection technique identifies samples of AGN missed by the other
technique.

Finally, using the greater statistics of the \xmm\, sample, we
consider if there are any trends between X-ray hardness ratio and
mid-infrared $W1 - W2$ color.  Though no strong correlation is
evident, we do find the expected general trend of redder mid-infrared
sources having harder X-ray spectra.  Splitting the X-ray sample
at $HR = 0$, the softer mid-infrared AGN candidates (\eg, $HR < 0$)
have a mean \wise\, color of $\langle W1 - W2 \rangle = 1.18$.
In contrast, the harder mid-infrared AGN candidates (\eg,
$HR > 0$) have a mean \wise\, color of $\langle W1 - W2 \rangle =
1.32$.  However, there are examples of very hard X-ray
sources with relatively blue \wise\, colors, as well as very soft
X-ray sources with relatively red \wise\, colors.


\scriptsize
\begin{deluxetable}{cccccccl}
\tablecaption{\wise$+$IRAC AGN candidates undetected by \xmm.}
\tablehead{
\colhead{\wise\, ID} &
\colhead{$i$} &
\colhead{$W1$} &
\colhead{$W2$} &
\colhead{$W1 - W2$} &
\colhead{[5.8]$-$[8.0]} &
\colhead{$z$} &
\colhead{Notes}}
\startdata
J095733.79+020943.1 & 19.66 & 16.33 & 15.08 & 1.25 & 1.07 &  1.441 & SDSS QSO \\
J095736.56+020236.7 & 21.31 & 16.12 & 14.74 & 1.38 & 1.35 &  \nodata& \\
J095752.32+022021.2 & 18.80 & 15.75 & 14.47 & 1.28 & 1.32 &  2.050 & SDSS QSO \\
J095756.65+020719.5 & 20.27 & 15.67 & 14.87 & 0.80 & 0.78 & [0.31] & \\
J095821.40+025259.0 & 19.53 & 15.67 & 14.50 & 1.17 & 1.29 & \nodata& \\
J095855.40+022037.4 & 19.43 & 15.41 & 14.59 & 0.82 & 1.39 & [0.38] & bright galaxy \\
J095905.55+025145.0 & 19.85 & 15.95 & 15.08 & 0.87 & 1.35 & \nodata& \\
J095937.35+021905.9 & 22.30 & 16.00 & 14.70 & 1.30 & 1.17 &  0.927 & \\
J095945.60+013032.2 & 20.34 & 15.97 & 15.01 & 0.97 & 0.99 &  1.106 & SDSS QSO \\
J100006.19+015535.3 & 20.90 & 15.53 & 14.19 & 1.34 & 1.05 &  0.661 & \\
J100008.42+020247.4 & 20.24 & 15.87 & 14.82 & 1.05 & 1.12 &  0.370 & type-2 AGN \\
J100008.93+021440.5 & 18.93 & 15.93 & 14.72 & 1.20 & 1.26 &  2.536 & QSO \\
J100013.51+013739.2 & 19.94 & 16.23 & 14.83 & 1.40 & 1.12 &  1.608 & SDSS QSO \\
J100043.70+014202.5 & 21.77 & 15.89 & 14.89 & 1.00 & 1.40 &  0.741 & \\
J100046.91+020726.5 & 21.86 & 14.87 & 13.14 & 1.73 & 1.18 &  1.158 & faint \\
J100115.38+024231.4 & 20.44 & 16.21 & 15.02 & 1.20 & 1.13 & \nodata& \\
J100135.61+022104.8 & 25.11 & 16.98 & 14.97 & 2.01 & 1.49 & \nodata& faint \\
J100137.11+024650.6 & 21.58 & 16.36 & 14.87 & 1.49 & 1.16 &  0.143 & \\
J100142.22+024330.7 & 22.69 & 15.74 & 14.02 & 1.72 & 1.28 & [1.62] & \\
J100231.86+015242.3 & 21.01 & 15.85 & 14.68 & 1.17 & 0.99 & \nodata& \\
J100244.77+025651.5 & 20.26 & 15.86 & 14.72 & 1.14 & 1.09 & \nodata& \\
J100246.35+024609.6 & 19.08 & 15.63 & 14.69 & 0.94 & 1.37 & \nodata& \\
J100253.31+020222.1 & 20.84 & 15.32 & 14.24 & 1.08 & 1.61 &  0.902 & \\
J100303.46+022632.0 & 20.69 & 15.68 & 14.82 & 0.85 & 1.14 & \nodata& \\
J100305.98+015704.0 & 19.46 & 14.63 & 12.91 & 1.72 & 1.32 &  0.370 & type-2 AGN \\
J100322.00+014356.5 &\nodata& 16.21 & 15.08 & 1.12 & 1.05 & \nodata& \\
%
\enddata
\label{table:xmm}
\tablecomments{Bracketed redshifts indicate photometric redshifts.}
\end{deluxetable}
\normalsize

%

\begin{figure}[t!]
\plotfiddle{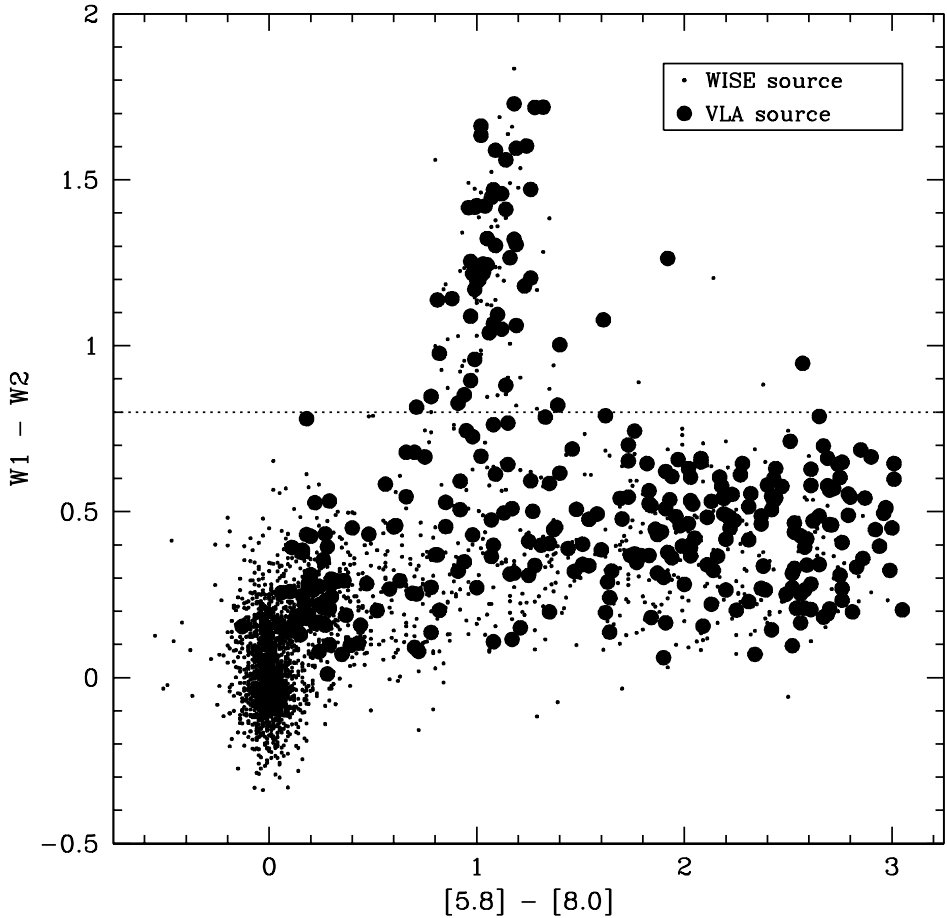}{4.5in}{0}{65}{65}{-210}{-105}
\caption{Mid-infrared color-color diagram of \wise-selected sources
in the COSMOS field, with \wise\, $W1-W2$ plotted against IRAC
$[5.8]-[8.0]$.  Plotted sources and symbols are as in
Fig.~\ref{fig:color_color1}, with large black circles added for all
\wise\, sources with VLA counterparts in the VLA-COSMOS survey.
\label{fig:color_colorVLA}}
\end{figure}

\subsection{Radio Properties of WISE AGN}

The Very Large Array (VLA) obtained deep radio images of the COSMOS
field at 20~cm.  The goals, observing strategy, and data reductions
for this large program, called the VLA-COSMOS survey, are described
in \citet{Schinnerer:07}.  The survey entailed nearly 350~hr of
exposure time, primarily in the highest resolution, or A, configuration.
The Large project imaged the full 2 deg$^2$ COSMOS field with a
resolution of 1\farcs5 to a sensitivity of $\sim 11 \mu$Jy (1$\sigma$)
\citep{Bondi:08}.  We use the joint catalog of \citet{Schinnerer:10},
which combines an improved analysis of the Large project with data
from the Deep project that doubled the integration time in the
central 0.84 deg$^2$ region of the survey.  The Joint catalog includes
2865 sources, of which 131 consist of multiple components.  

We match the full Joint catalog to the full \wise\, source list of
3618 sources using a 1\farcs5 matching radius.  We find 333 matches,
of which 33 have multiple components in the VLA data.
Fig.~\ref{fig:color_colorVLA} shows the mid-infrared colors of the
VLA sources, which shows many radio-detected sources on both sides
of our $W1 - W2 = 0.8$ color cut.  This is unsurprising, as these
extremely deep radio data detect emission related to stellar
processes (e.g., supernova remnants) as well as AGN activity at a
range of Eddington ratios.  Considering the \wise-selected AGN
candidates, 55 of the 130 candidates (42\%) have radio matches;
five of these are flagged as consisting of multiple components in
the radio data.  Of the 123 AGN candidates identified by both \wise\,
and IRAC, 52 (42\%) are detected by these very deep radio data.
For the \wise-selected AGN not flagged as likely AGN by IRAC, 
three (43\%) are detected by the VLA.  All of these fractions are
much higher than the 8\% of sources with $W1 - W2 < 0.8$ which are
detected by the VLA-COSMOS survey.

Assuming a typical quasar radio spectral index $\alpha = -0.5$
($S_\nu \propto \nu^\alpha$) and adopting the \citet{Gregg:96}
cutoff value for the 1.4~GHz specific luminosity, $L_{\rm 1.4 GHz}
= 10^{32.5}\, h_{50}^{-2}\, {\rm erg}\, {\rm s}^{-1}\, {\rm Hz}^{-1}$
($\approx 10^{24}\, h_{50}^{-2}\, {\rm W}\, {\rm Hz}^{-1}\, {\rm
sr}^{-1}$), to discriminate radio-loud and AGN radio-quiet populations,
only two of 98 the \wise-selected AGN with spectroscopic redshifts
are radio-loud\footnote{For consistency with previous work in terms
of defining the boundary between radio-loud and radio-quiet
populations, 1.4~GHz specific luminosity is calculated for an
Einstein-de~Sitter cosmology; e.g., see \citet{Stern:00a}.}.  The results
are unchanged if we assume $\alpha = -0.8$, as might be more typical for
quasars without the jet aligned along our line of sight.  Both
of these sources are SDSS quasars at $z > 1$ (WISE~J095821.65+024628.2
at $z = 1.405$ and WISE~J095908.32+024309.6 at $z = 1.318$).  These
two sources come from a total of 45 \wise-selected AGN that are
optically bright and have spectroscopy from the SDSS; all are
classified as broad-lined, and 35 are at $z > 1$ implying that they
are clearly luminous quasars.  Considering just this SDSS subsample
of \wise-selected AGN, our results are statistically consistent
with the canonical value of $\sim 10\%$ of quasars being radio-loud
\citep[\eg,][]{Stern:00a}.  However, the fact that none of the 53
other sources with spectroscopic redshifts are radio-loud is
surprising, suggesting that the radio-loud fraction might be different
for type~2 AGN.  We note, however, that \citet{Zakamska:04} found
no change in the radio-loud fraction for their sample of SDSS-selected
obscured quasars.

Note that radio-loud AGN do not fall below the 5$\sigma$ VLA-COSMOS
sensitivity until beyond $z \sim 10$; e.g., the survey is sensitive
enough to detect all radio-loud AGN that \wise\, is likely to detect.
All of the \wise-selected AGN candidates without redshifts that
were detected by the VLA-COSMOS survey have $S_\nu < 0.72$~mJy,
implying that they would have to be at $z > 3.4$ in order to be
radio-loud.

\begin{figure}[t!]
\plotfiddle{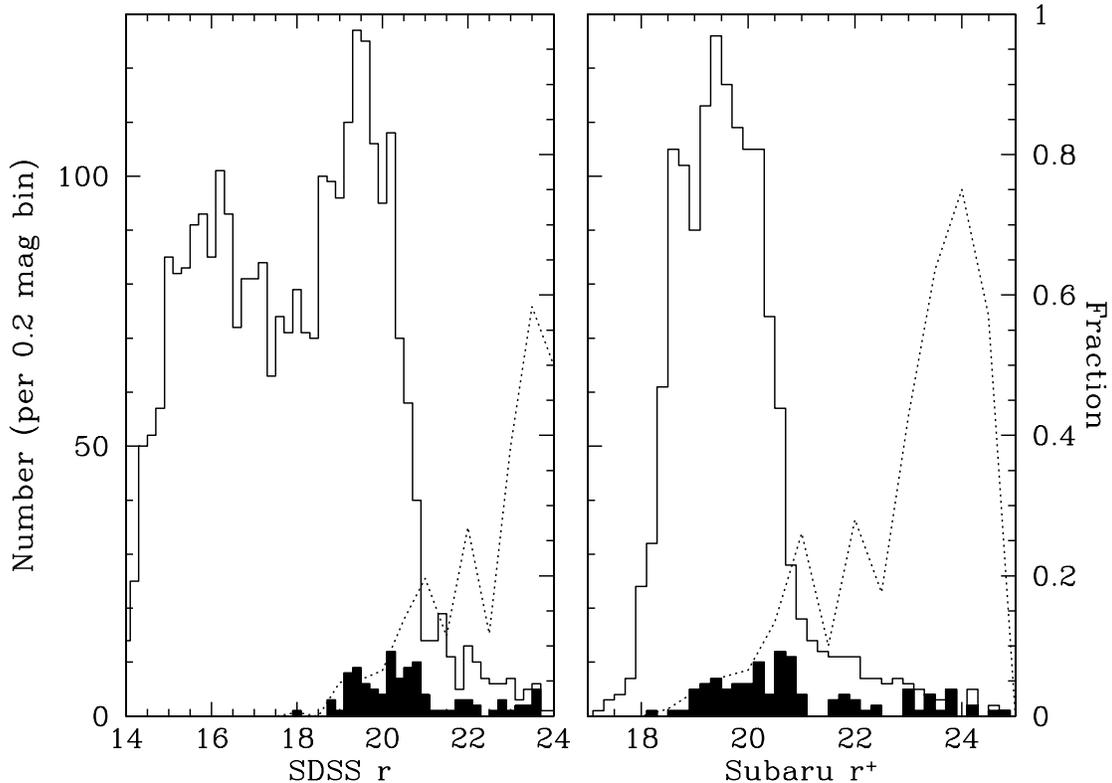}{3.8in}{-90}{55}{55}{-210}{315}
\caption{Histograms of SDSS $r$-band (left) and Subaru $r^+$-band
(right) optical magnitudes of all \wise-selected sources in the COSMOS
field (open histogram).  Solid histograms show distributions for
\wise-selected AGN candidates.  Dotted lines show the fraction of
\wise\, sources that are AGN candidates as a function of optical
magnitude (in 0.5 mag bins).  While very few sources brighter than
$\sim 19$th mag are AGN candidates, the fraction increases to $>
50$\% at the faintest magnitudes.
\label{fig:rhist}}
\end{figure}

\begin{figure}[t!]
\plotfiddle{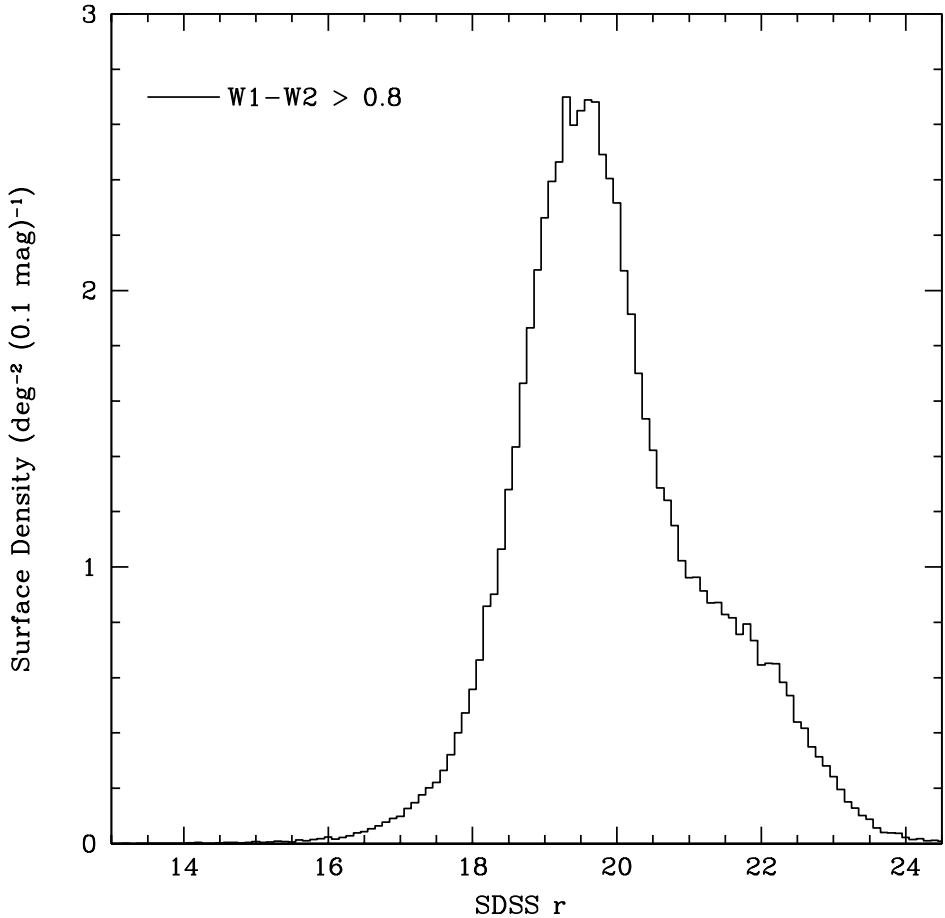}{4.5in}{0}{65}{65}{-210}{-105}
\caption{Distribution of $r$-band magnitudes for \wise-selected AGN
candidates identified over 4000 deg$^2$ of the SDSS.
\label{fig:sdssrhist}}
\end{figure}

\begin{figure}[t!]
\plotfiddle{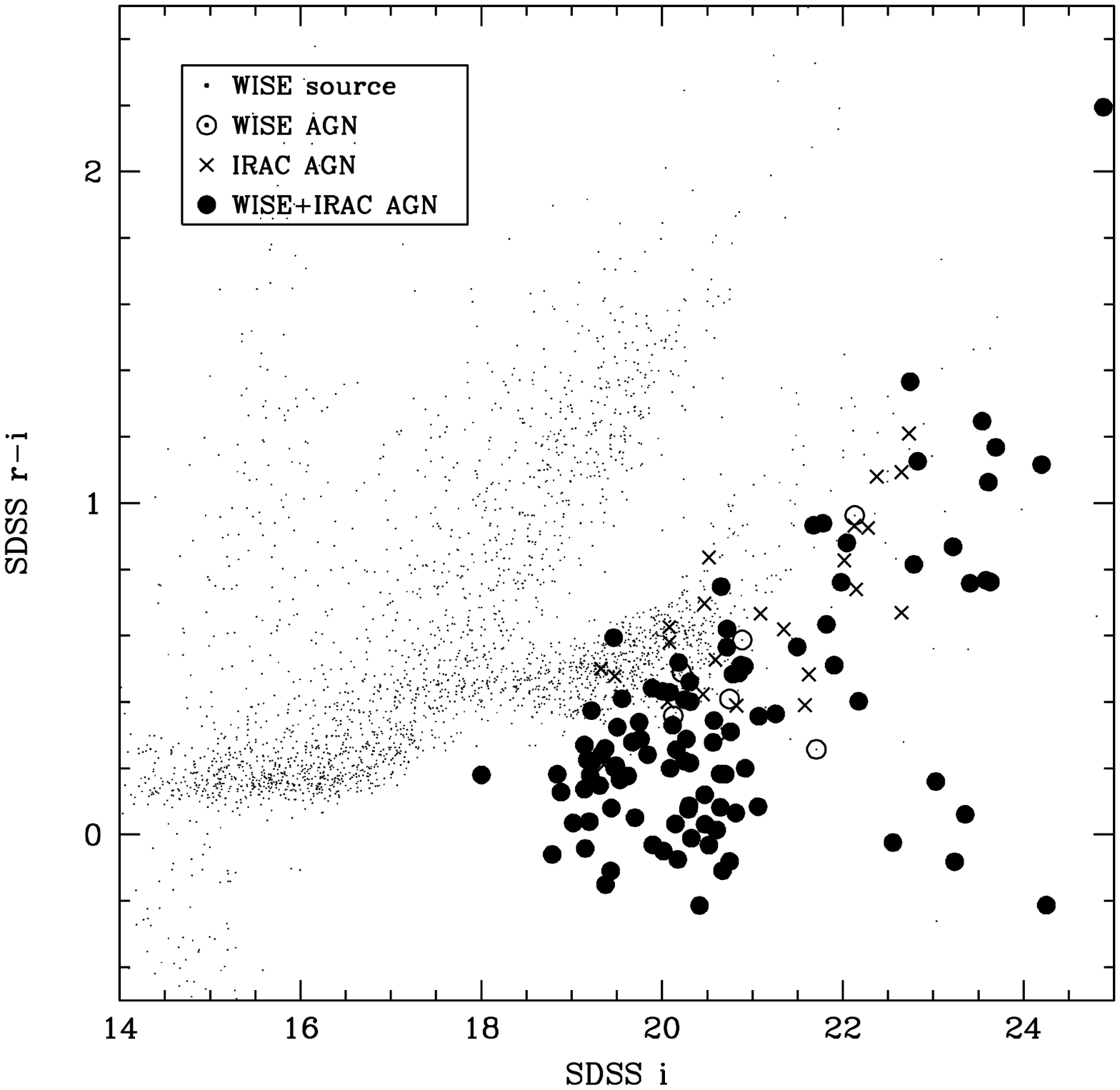}{4.5in}{0}{65}{65}{-210}{-105}
\caption{Optical color-magnitude diagram of \wise\, sources in the
COSMOS field.  Optical photometry is from SDSS, and symbols are
indicated in the upper left.  \wise-selected AGN candidates tend
to optical colors that are bluer than typical field sources, though
a significant fraction of the AGN candidates overlap with the field
population.
\label{fig:sdss}}
\end{figure}

\subsection{Optical Magnitudes and Colors of WISE AGN}

We next consider the optical properties of the \wise-selected AGN
candidates.  Fig.~\ref{fig:rhist} shows the $r$-band magnitude
distributions of all \wise\, sources in the COSMOS field (open
histogram) as well as the AGN candidates with $W1 - W2 \geq 0.8$
(solid histogram).  All photometry is in the AB system and comes
from the COSMOS photometry catalog of \citet{Capak:07}, which is
available through IRSA.  The left panel shows $r$-band photometry
from the second data release (DR2) of the SDSS \citep{Abazajian:04}.
Objects as bright as 10th mag have good photometry in the SDSS
imaging, and the imaging depth, defined as the 95\% completeness
limit for point sources, is $r \sim 22.2$.  Similar data is available
over more than 11,000 deg$^2$.  However, many of the \wise\, sources
are fainter than this limiting depth, so in the right panel of
Fig.~\ref{fig:rhist} we show the Subaru $r^+$ photometry in the
COSMOS field obtained with the Suprime-Cam instrument \citep{Komiyama:03}.
These data reach a 5$\sigma$ depth (3\arcsec\ aperture) of 26.6 and
detect all of the \wise-selected AGN candidates.  Many of the
brighter sources in the Subaru data are unresolved, leading to
saturation issues.  This causes the truncation seen at $r^+ \simlt
18$.

Fig.~\ref{fig:rhist} shows that very few of the \wise-selected AGN
candidates are brighter than $r \sim 18$, and they represent less
than 5\% of the \wise\, source population at bright optical magnitudes.
However, the AGN candidates represent an increasing fraction of the
optically fainter \wise\, sources, accounting for $\sim 20$\% of
\wise\, sources at $r \sim 21$ and more than 50\% of \wise\, sources
with $r \simgt 23$.  In order to explore the optical brightnesses
of \wise-selected AGN candidates with higher fidelity, we identified
AGN candidates over 4000 deg$^2$ of the SDSS \citep[c.f.,][]{Donoso:12,
Yan:12}.  Fig.~\ref{fig:sdssrhist} shows the resultant $r$-band
distribution.  While the majority of AGN candidates are well detected
in the SDSS imaging, their optical brightness distribution peaks
at $r \sim 19.5$, making them fainter than the typical spectroscopic
limits of SDSS, $i \sim 19.1$ for quasars and $r \sim 17.8$ for
galaxies (as discussed in the following section, approximately half
of our \wise-selected AGN candidates are spatially resolved).

Fig.~\ref{fig:sdss} shows an optical color-magnitude diagram of \wise\,
sources.  At bright magnitudes, the distribution is dominated by
Galactic stars with $r - i \sim 0.1$, while a second galaxy sequence
becomes evident at $r \simgt 18$.  Mid-infrared AGN candidates are, on
average, bluer than typical galaxies, though the color distribution
clearly overlaps with the galaxy sample.  AGN candidates identified
by both \wise\, and {\it Spitzer}/IRAC, which represent the most
robust AGN sample with the highest rate of X-ray detections, tend
to be bluer than galaxies at $i \simlt 21$, though at fainter
magnitudes where obscured AGN become more prevalent, the distribution
fans out.  This is partially due to photometric errors, but also
because a large fraction of the AGN candidates have colors similar
to galaxies.  Not surprisingly, AGN candidates identified by {\it
Spitzer} but not by \wise --- \eg, sources at the blue corner of the
\citet{Stern:05b} wedge and close to the galaxy locus in mid-infrared
color-color space --- tend to have galaxy-like optical colors.

\begin{figure}[t!]
\plotfiddle{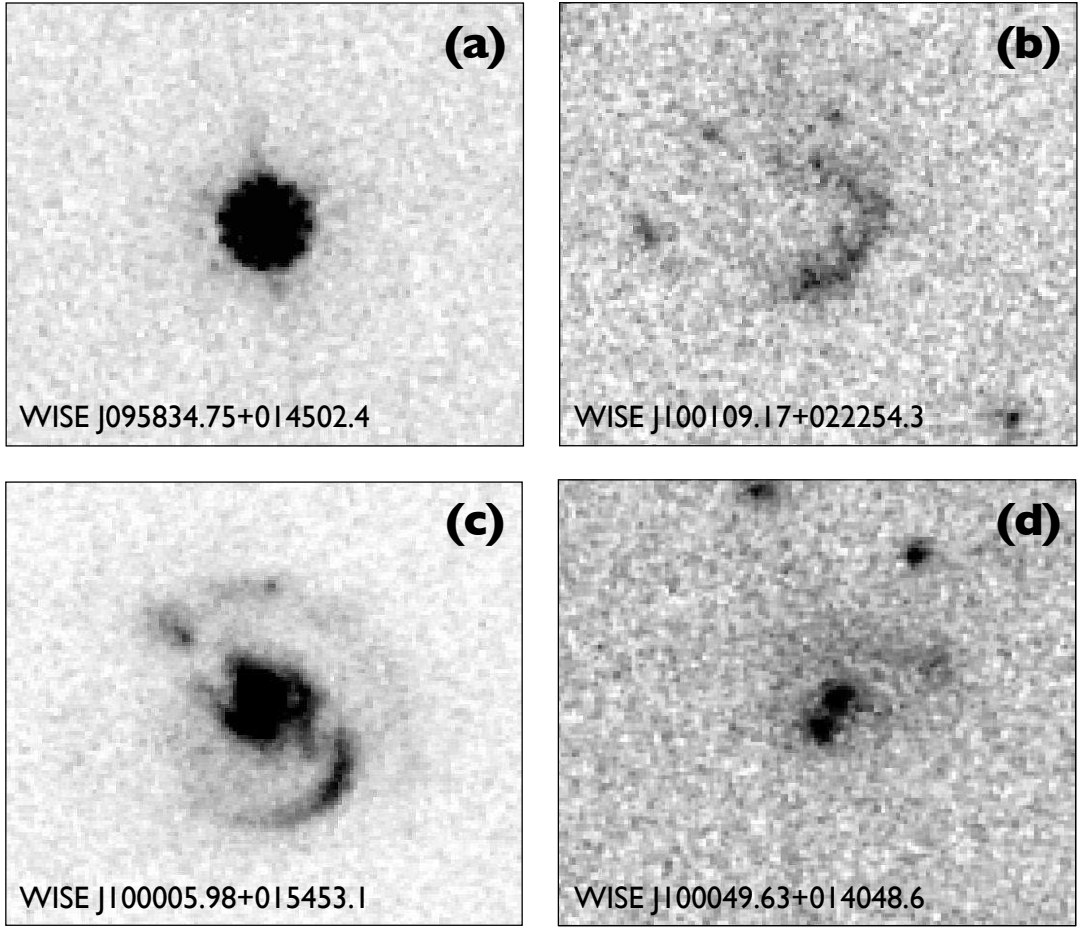}{4.7in}{90}{65}{65}{260}{-20}
\caption{{\it Hubble}/ACS $I_{814}$ images of four \wise-selected
AGN in the COSMOS field, showing the range of optical morphologies.
Approximately half of the sources are unresolved point sources
(e.g., panel a).  The other half are spatially resolved (e.g.,
panels b-d), sometimes with rather faint optical magnitudes.  All
four sources shown here are identified as candidate AGN by both
\wise\, and IRAC color criteria.  Images are $\sim 4\arcsec$ on a
side (\eg, smaller than the \wise\, PSF), with North up and East
to the left.
\label{fig:acs}}
\end{figure}

\subsection{Morphologies of WISE AGN}

Conventional wisdom states that the most luminous AGN in the universe
are associated with unresolved point sources at optical wavelengths.
While this is true for the vast majority of unobscured, type~1
quasars, this is not the case for obscured, type~2 quasars.  For
instance, luminous high-redshift radio galaxies often have a clumpy,
irregular morphology at rest-frame ultraviolet wavelengths, with
the emission generally elongated and aligned with the radio source
axis \citep[\eg,][]{McCarthy:87}.  At rest-frame optical wavelengths,
where stars dominate the galaxy luminosity, the hosts of most
luminous radio galaxies are normal elliptical galaxies with
$r^{1/4}$-law light profiles \citep[\eg,][]{Zirm:03}.

The cornerstone data set for the COSMOS survey is its wide-field
{\it Hubble Space Telescope} Advanced Camera for Surveys (ACS)
imaging \citep{Scoville:07, Koekemoer:07}.  With 583 single-orbit
F814W ($I_{814}$ hereafter) observations, these data cover (or
define) the 1.8 deg$^2$ COSMOS field and constitute the largest
contiguous {\it Hubble} imaging survey to date.  The data are
extremely sensitive, with 0\farcs09 resolution (FWHM) and reaching
a 50\% completeness limit of $I_{814} = 26.0$ (AB) for sources
0\farcs5 in diameter.

\citet{Griffith:10} have recently analyzed the optical morphologies
of AGN in the COSMOS field identified from a variety of methods:
radio selection, X-ray selection, and mid-infrared
selection \citep[see also][]{Gabor:09}.  They find that the
radio-selected AGN are likely to be hosted by early-type galaxies,
while X-ray and mid-infrared selected AGN are more often associated
with point sources and disk galaxies.  Considering just the brighter
X-ray and mid-infrared subsamples, approximately half of the AGN
are optically unresolved and a third are associated with disk
galaxies.  These morphological results conform with the results of
\citet{Hickox:09} who studied the colors and large-scale clustering
of AGN, and found a general association of radio-selected AGN with
``red sequence'' galaxies \citep[an old, well-known result;
\eg,][]{Matthews:64}, mid-infrared selected AGN are associated with
``blue cloud'' galaxies, and X-ray selected AGN straddle these
samples in the ``green valley''.

We find similar results here.  Of the 130 \wise-selected AGN
candidates, 94 are located within the portion of COSMOS imaged by
ACS.   A bit more than half (52/94, or 55\%) of the sources are
spatially resolved; the other AGN candidates are associated with
point sources.  As an aside, we note that one of the contaminants,
WISE~J100050.63+024901.7 was flagged by \citet{Faure:08} as one
of the 20 most likely strong lensing systems in the COSMOS field
\citep[see also][]{Jackson:08}.  The ACS $I_{814}$ image of this
system shows four faint arcs surrounding a bright early-type galaxy,
with a radial separation of 1\farcs9.  Inspection of the IRAC images
for this system shows that the mid-infrared data are still dominated
by the optically bright lensing galaxy.

Fig.~\ref{fig:acs} shows {\it Hubble}/ACS $I_{814}$ images of four
example \wise-selected AGN.  All four examples were also identified
as AGN candidates by their IRAC colors.  Several are X-ray and/or
radio sources as well.  Panel~(a) shows WISE~J095834.75+014502.4,
a bright SDSS quasar at $z = 1.889$.  It is unresolved in the ACS
image; approximately half of the \wise\, AGN candidates have similar
morphologies.  Panel~(b) shows WISE~J100109.21+022254.2, one of the
optically faintest \wise-selected AGN candidates in the COSMOS
field, with $I_{814} = 22.9$.  As discussed in the next section,
we were unable to obtain a redshift for this source from deep Keck
spectroscopy, though subsequent to our observations, \citet{Brusa:10}
reported that this source is as a narrow-lined AGN at $z = 1.582$.
Panels~(c) and (d) show two $z \sim 0.9$ type~2 (e.g., narrow-lined)
AGN from \citet{Trump:07}.  The former is WISE~J100005.99+015453.3,
which is a spiral galaxy with a very bright nucleus.  The latter
is WISE~J100013.42+021400.4, which has a more irregular morphology.

\begin{figure}[t!]
\plotfiddle{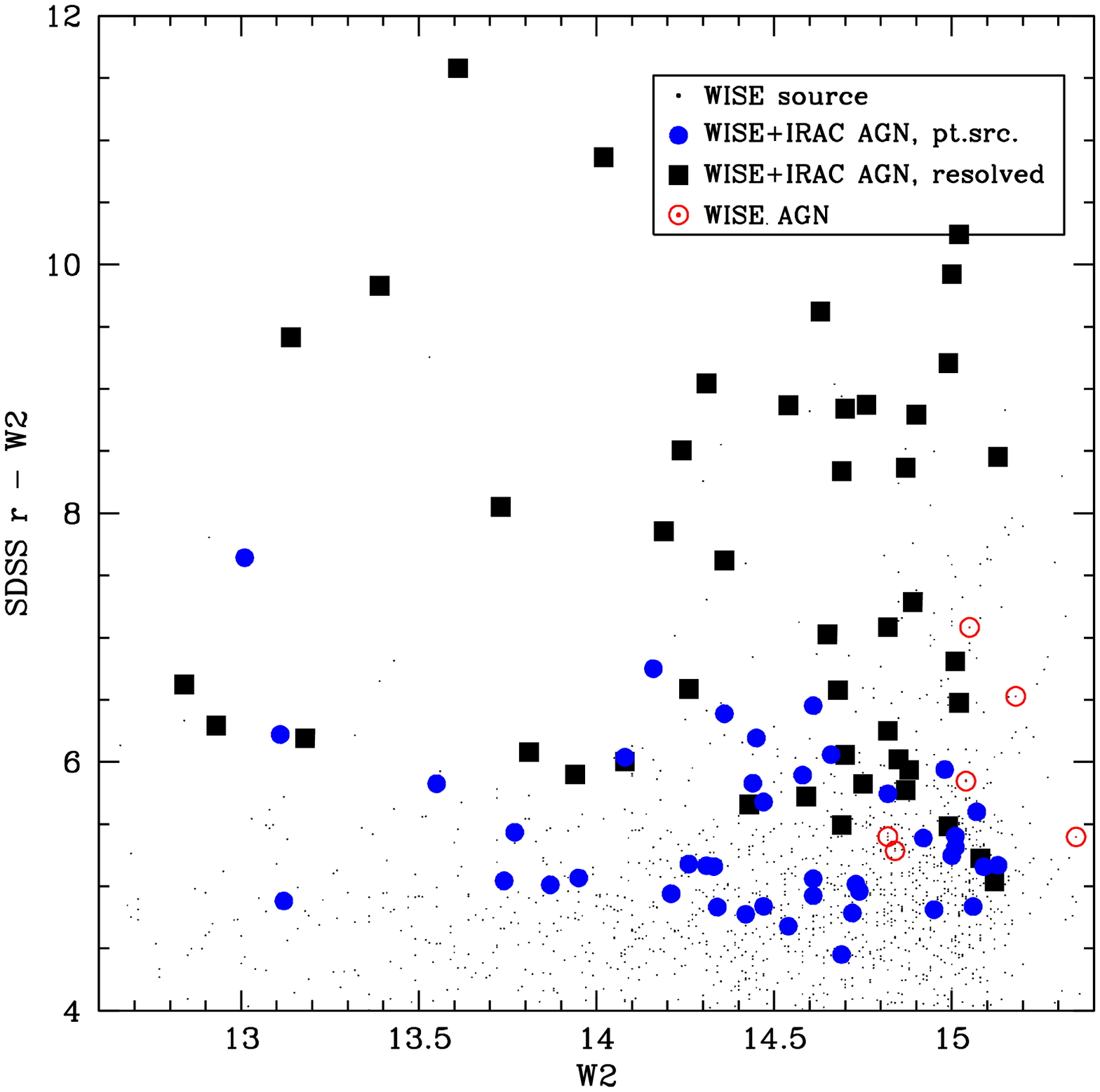}{4.5in}{0}{65}{65}{-210}{-105}
\caption{Optical-to-mid-infrared vs. mid-infrared color-magnitude
diagram of \wise-selected sources in the COSMOS field.  Optical
photometry is from SDSS, and symbols are indicated in the upper
left.  Only those AGN candidates covered by the COSMOS {\it Hubble}
images are plotted, and they are flagged by their ACS morphologies.
\label{fig:morph}}
\end{figure}

Fig.~\ref{fig:morph} presents a color-magnitude diagram of \wise\,
sources in the COSMOS field, with optical-to-mid-infrared color
plotted against 4.6~$\mu$m brightness ($W2$).  For the AGN candidates,
we only plot the $\sim 70\%$ of sources covered by the ACS imaging.
There are several things to note from this plot.  First, the
\wise-only AGN candidates (\eg, \wise-selected AGN candidates not
identified as AGN candidates from their \spitzer\, colors) clearly
reside on the right side of Fig.~\ref{fig:morph}, with no contaminants
brighter than $W2 = 14.8$.  This implies that caution should be
applied before extending our simple \wise\, color criterion to
fainter limits.  Indeed, in \citet{Assef:12} we investigate the
interloper fraction as a function of $W2$ magnitude and derive a
magnitude-dependent \wise\, AGN selection criterion applicable to
higher ecliptic latitude (\eg, deeper) portions of the \wise\,
survey.

Second, optically unresolved AGN candidates tend to have bluer $r
- W2$ colors, consistent with the expectation that they suffer less
extinction at optical wavelengths.  Quantitatively, the unresolved
\wise+IRAC AGN candidates in Fig.~\ref{fig:morph} have $\langle r
- W2 \rangle = 5.18$.  Similarly selected sources that are resolved
in the {\it Hubble} imaging have $\langle r - W2 \rangle = 7.29$.
The unresolved AGN candidates are also slightly brighter, with
$\langle W2 \rangle = 14.54$ as compared to $\langle W2 \rangle =
14.69$ for the resolved \wise+IRAC AGN candidates.  Importantly,
however, we note that both resolved and unresolved sources are found
across the full $W2$ range probed.

Finally, we also consider the optical properties of the \wise+IRAC
AGN candidates that are undetected by \xmm\, (Table~\ref{table:xmm}).
These sources have an even fainter median mid-IR brightness, $\langle
W2 \rangle = 14.72$, and also have redder optical-to-mid-IR colors
than typical \wise+IRAC AGN candidates.  Quantitatively,  $\langle
r - W2 \rangle = 6.58$ for the X-ray-undetected AGN candidates,
while  $\langle r - W2 \rangle = 5.85$ for the X-ray detected
\wise+IRAC AGN sample.  This is consistent with the X-ray undetected
sources being associated with more heavily obscured AGN, diminishing
both their optical and X-ray fluxes.

\subsection{Redshift Distribution of WISE AGN}

In order to understand the redshift distribution and properties of
\wise-selected AGN candidates, we have both matched the candidate
list to publically available spectroscopy in the COSMOS field and
obtained new observations.  Published spectroscopy come from several
papers:  bright targets have spectroscopic redshifts from the SDSS
\citep{Abazajian:09}; \citet{Prescott:06} reports on MMT/Hectospec
follow-up of optically selected quasar candidates in COSMOS;
\citet{Lilly:07} reports on zCOSMOS, a large VLT/VIMOS $I$-band
magnitude-limited survey of the COSMOS field; \citet{Trump:07} and
\citet{Trump:09} report on Magellan/IMACS spectroscopy of X-ray-
and radio-selected AGN candidates in the COSMOS field; and
\citet{Brusa:10} report on spectroscopy of X-ray sources from the
{\it XMM-Newton} wide-field survey of the COSMOS field, synthesizing
both previously published results and new spectroscopy from Keck.


We obtained additional spectroscopy on UT 2010 March 12-15 using
the Low Resolution Imaging Spectrometer \citep[LRIS;][]{Oke:95} and
the DEep Imaging Multi-Object Spectrograph \citep[DEIMOS;][]{Faber:03}.
We observed three Keck slitmasks in the COSMOS field.  On UT 2010
March 12 we observed {\tt cos10b} for 5200~s using the dual-beam
LRIS instrument.  We used the 400 $\ell$ mm$^{-1}$ grism on the
blue arm of the spectrograph (blazed at 3400~\AA; resolving power
$R \equiv \lambda / \Delta \lambda \sim 600$), the 400 $\ell$
mm$^{-1}$ grating on the red arm of the spectrograph (blazed at
8500~\AA; $R \sim 700$) and 6800~\AA\ dichroic.  On UT 2010 March
13 we observed {\tt cos10a} for 1200~s using LRIS.  The red CCD was
non-functional that night, so we channeled all of the light to the
blue arm of the spectrograph and again used the 400 $\ell$ mm$^{-1}$
grism blazed at 3400~\AA.  On UT 2010 March 14 we observed {\tt
cos10d} for 3600~s with DEIMOS in cloudy conditions, using the 600
$\ell$ mm$^{-1}$ grating (blazed at 7500~\AA; $R \sim 1600$) and
the 4000~\AA\ order-blocking filter.  Masks all used $\sim 1\farcs2$
wide slitlets.  Data reduction followed standard procedures, and
we flux calibrated the data using observations of standard stars
from \citet{Massey:90}.  Note that the DEIMOS data were taken in
non-photometric conditions, resulting in an uncertainty in the flux
scale of those sources.

\begin{figure}[t!]
\plotfiddle{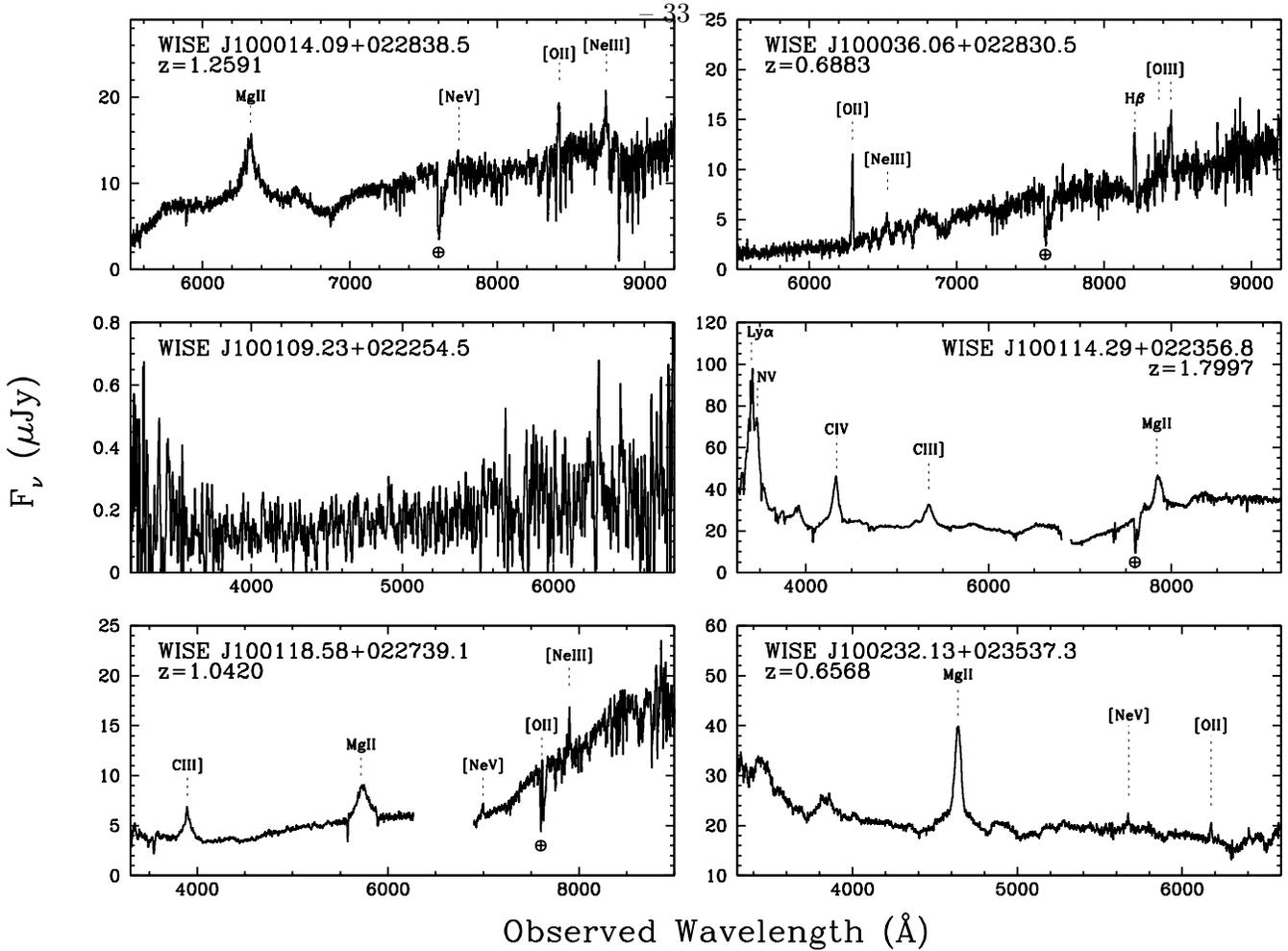}{4.5in}{-90}{70}{70}{-280}{400}
\caption{Results from Keck spectroscopy of sources identified as
AGN candidates based on their \wise\, colors, obtained in March
2010.  All six sources here are also selected as AGN candidates
based on their {\it Spitzer}/IRAC colors (Stern \etal\ 2005).
Prominent emission lines are marked, as is the telluric A-band
absorption at 7600 \AA.  DEIMOS spectra (see Table~3) were obtained
in non-photometric conditions; the relative calibration of such
sources should be reliable, though the absolute scale is uncertain.
\label{fig:keck}}
\end{figure}

Target selection was done prior to access to the \wise\ data in the
COSMOS field, though we had already anticipated that red $W1 - W2$
colors would be an effective method to identify a large population
of AGN.  We sought to test that hypothesis using {\it Spitzer}/IRAC
imaging from the S-COSMOS survey \citep{Sanders:07}, assuming that
$W1 \sim$ [3.6] and $W2 \sim$ [4.5].  Additional targets were
selected using the \citet{Stern:05b} IRAC AGN wedge selection
criteria.  Fig.~\ref{fig:keck} and Table~\ref{table:keck} present
the results for the six COSMOS targets that subsequently were found
to match our $W1 - W2 \geq 0.8$ AGN candidate selection criterion.
All six would also be selected by the \citet{Stern:05b} IRAC
criteria.  Table~\ref{table:appendix} in the Appendix presents the
results for the additional COSMOS targets observed on these masks.
Our new Keck results are occasionally slightly discrepant with
previous results, but typically with $\Delta z \leq 0.01$.  The
signal-to-noise ratio of these new data are quite high and the data
were taken at relatively high spectral dispersion, suggesting that
these redshifts should take precedence over previous results.

Four of the sources show prominent AGN features, such as broadened
\ion{Mg}{2}~2800 emission.  WISE~J100036.06+022830.5 does not show
obvious AGN features; the spectrum shows narrow emission lines from
[\ion{O}{2}], [\ion{Ne}{3}], H$\beta$, and [\ion{O}{3}], as well
as Balmer absorption lines indicative of a relatively young stellar
population.  We find $\log$ ([\ion{O}{3}] / H$\beta$) $\sim 0.12$,
which is consistent with both star-forming and AGN activity in the
\citet{Baldwin:81} diagram; spectral features redward of our data
are required to distinguish the principle line excitation mechanism.
We did not obtain a redshift for WISE~J100109.23+022254.5 with our
data, though \citet{Brusa:10} identify this source as a narrow-lined
AGN $z = 1.582$ on the basis of their deep Keck/DEIMOS spectroscopy;
the quality of the redshift is not indicated.  Our spectroscopy
does not show any features such as redshifted \ion{C}{4} emission
or absorption to confirm that redshift.  However, we note that
strongest feature at this redshift is likely to be [\ion{O}{2}]
emission at 9623 \AA.  This is beyond the wavelength coverage of
our Keck spectroscopy.

\begin{figure}[t!]
\plotfiddle{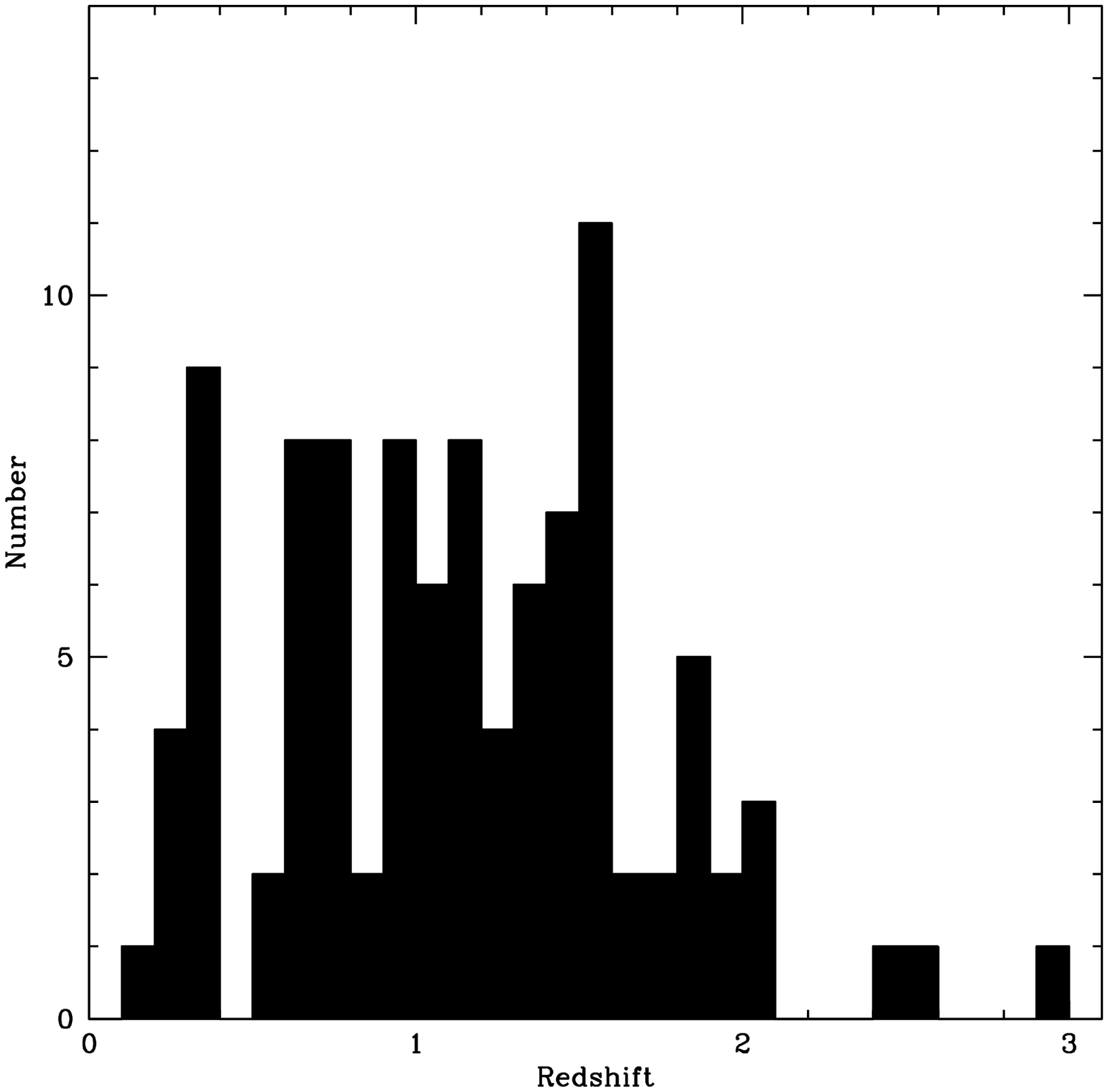}{4.5in}{0}{65}{65}{-210}{-105}
\caption{Histogram of spectroscopic redshifts for \wise-selected
AGN candidates in the COSMOS field.  Of the 130 such candidates,
101 have spectroscopic redshifts. The median redshift is $\langle
z \rangle = 1.11$.
\label{fig:zhist}}
\end{figure}

Fig.~\ref{fig:zhist} presents the distribution of spectroscopic
redshifts for the \wise\, AGN candidates.  We have spectroscopic
redshifts for 101 of the 130 candidates (72\%); the median redshift
is $\langle z \rangle = 1.11$.  Seven of these candidates are from
outside the IRAC AGN wedge, four of which have spectroscopic
redshifts.  Two are broad-lined quasars at $z \sim 1$; the other
two are galaxies at $z = 0.27$ and $z = 0.75$ from zCOSMOS.  This
suggests that the 95\% reliability rate derived in \S~3.1 is actually
a lower limit; some of the \wise-selected candidates are indeed AGN
despite not being identified as such by their IRAC colors.

\scriptsize
\begin{deluxetable}{cccccl}
\tablecaption{Results of 2010 March Keck Observations.}
\tablehead{
\colhead{$I_{814}$} &
\colhead{R.A.} &
\colhead{Dec.} &
\colhead{$z$} &
\colhead{Slitmask(s)} &
\colhead{Notes}}
\startdata
20.52 & 10:00:14.09 & +02:28:38.5 & 1.2591 & D[6]   & QSO: MgII,[OII],[NeIII] \\ 						
20.47 & 10:00:36.06 & +02:28:30.5 & 0.6883 & D[5]   & [OII],H$\beta$,[OIII] \\						
22.91 & 10:01:09.23 & +02:22:54.5 &        & B,D[2] & faint blue cont.; $z=1.582$ in Brusa \etal\ (2010) \\
19.16 & 10:01:14.29 & +02:23:56.8 & 1.7997 & B,D[1] & QSO: Ly$\alpha$,CIV,CIII],MgII \\					
20.04 & 10:01:18.58 & +02:27:39.1 & 1.0420 & B      & QSO: CIV,CIII],MgII,[NeV],[OII],[NeIII] \\    				
18.85 & 10:02:32.13 & +02:35:37.3 & 0.6568 & A      & QSO: MgII,[NeIV],[OII]; jet? \\					
19.30 & 10:00:22.79 & +02:25:30.6 & 0.3482 & D[3]   & H$\alpha$ \\ 										
\enddata
\label{table:keck}

\tablecomments{Masks A and B were observed with LRIS.  Mask D was
observed with DEIMOS; the bracketed numbers indicate the DEIMOS
slitlet number.  All derived redshifts are of very high quality (Q
= A; see Appendix~A).}

\end{deluxetable}
\normalsize

\section{Conclusions}


We use the deep, public, multiwavelength data in the $\sim 2$~deg$^2$
COSMOS survey to motivate a very simple, empirical mid-infrared
criterion to identify AGN candidates with the \wise\, satellite.
Selecting sources with $W1 - W2 \geq 0.8$ identifies $61.9 \pm 5.4$
AGN candidates per deg$^2$ at the $10 \sigma$ depth of the \wise\,
COSMOS data (\eg, 160 $\mu$Jy at 4.6 $\mu$m).  Using deep {\it
Spitzer} data in this field and adopting the mid-infrared two-color
AGN selection criteria of \citet{Stern:05b} as the truth sample,
this simple \wise\, color cut is approximately 78\% complete and
95\% reliable at identifying AGN.  Of the seven `contaminants' in
the COSMOS field identified as AGN candidates using our new \wise\,
color criterion but not selected as an AGN candidate from the
\spitzer\, color criteria, two are identified as broad-lined quasars,
implying that the reliability of this simple color selection is
better than 95\% at the depth of the \wise\, COSMOS observations,
$W2 \sim 15.0$.  We caution, however, that COSMOS, by design, is
at very low ecliptic latitude implying that its \wise\, coverage
is shallower than average.  In deeper \wise\, fields, 
this simple color cut suffers more contamination.
Here we show that the combined criteria $W1 - W2 \geq 0.8$ and $W2
\leq 15$ robustly identifies an extremely robust, highly complete
AGN sample.  In \citet{Assef:12}, we use the deeper, wider-area
Bo\"otes field to derive a $W2$-dependent AGN color selection
criterion that is applicable in deeper areas of the \wise\, all-sky
survey.


Forty-six of the \wise-selected AGN candidates in the COSMOS field
are known broad-lined quasars previously identified by the SDSS
(\eg, 21.9 type-1 quasars per deg$^{2}$).  The median optical-to-mid-IR
color of these type-1 AGN is $\langle i - W2 \rangle = 4.75$.  The
$10\sigma$ $W2$ depth that we applied to the \wise\, COSMOS
observations corresponds to $W2 = 15.05$, implying that our mid-IR
\wise\, AGN selection should identify unobscured quasars to an
optical depth of $i \sim 19.8$.  \citet{Richards:06a} combines the
SDSS and 2QZ/6QZ quasar surveys to study the demographics and
evolution of quasars below the SDSS photometric limits.  They find
$\sim 20$ type-1 quasars per deg$^{2}$ to this depth.  Assuming that
the other $\sim 40$ \wise-selected AGN candidates per deg$^{2}$ are
type-2 quasars, the implied obscured-to-unobscured ratio is $\sim
2:1$ at these bright depths.  This result is in-line with expected
ratios required to explain the intensity and hardness of the cosmic
X-ray background \citep[\eg,][]{Treister:04, Treister:05, Gilli:07,
Ballantyne:11}.


All of the \wise-selected AGN candidates in COSMOS have optical
identifications.  Approximately half are spatially resolved.
\wise-selected AGN tend to be amongst the optically faintest \wise\,
sources, accounting for essentially none of the \wise\, sources
brighter than $r = 18$, $\sim 20$\% of \wise\, sources at $r = 21$,
and rising to $> 50\%$ of sources fainter than $r = 23$.  The
$r$-band distribution of \wise-selected AGN candidates peaks at $r
\sim 19.5$, but has a significant tail to fainter magnitudes.
Considering the 101 candidates with spectroscopic redshifts, the
median redshift is $\langle z \rangle = 1.11$.

Most ($\sim 75 \%$) of the robust \wise\, AGN candidates covered
by the deep {\it Chandra} and {\it XMM-Newton} imaging of COSMOS
are detected at X-ray energies, while few of the expected contaminants
are.  Of particular note is the $\sim 25\%$ of robust AGN candidates
identified in 90~s \wise\, full-sky images that are missed in
extremely deep, 60+~ks pencil-beam surveys by these flagship-class
soft ($\simlt 10$~keV) X-ray missions.  Such sources are expected
to be heavily obscured, luminous, Compton-thick AGN.  In the next
year, the {\it NuSTAR} satellite will map the COSMOS field in the
$5 - 80$~keV hard X-ray energy range, reaching depths $\sim 200$
more sensitive than previous surveys in this energy range.  We
expect that several of the obscured \wise\, AGN candidates will be
detected by {\it NuSTAR}.


The 130 \wise-selected AGN candidates identified in the COSMOS field
is sufficiently large to characterize general properties of the
population, and the expectation is that this selection criterion
will be valuable for a wide range of future studies, such as
understanding the energetics of sources identified at other wavelengths
\citep[\eg,][]{Bond:12}, comparing the environments of type~1 and
type~2 AGN, and probing the role of AGN in galaxy formation and
evolution.  A companion paper, \cite{Assef:12}, uses nearly an
order of magnitude larger sample of \wise-selected AGN in the $\sim
10$ deg$^2$ Bo\"otes field to study the evolutionary properties of
this population.

\acknowledgements 

We gratefully acknowledge the anonymous referee for helpful comments
that have made the paper both clearer and stronger.  We also thank
P.~Capak for providing two unpublished redshifts obtained with
DEIMOS.  This publication makes use of data products from the {\it
Wide-field Infrared Survey Explorer}, which is a joint project of
the University of California, Los Angeles, and the Jet Propulsion
Laboratory/California Institute of Technology, funded by the National
Aeronautics and Space Administration.  We gratefully acknowledge
the COSMOS survey, and are thankful for the extensive and high
quality data products that they have publicly released.  This
publication makes use of data obtained at Keck Observatory.  The
authors wish to recognize and acknowledge the very significant
cultural role and reverence that the summit of Mauna Kea has always
had within the indigenous Hawaiian community; we are most fortunate
to have the opportunity to conduct observations from this mountain.
SDSS is funded by the Alfred P.  Sloan Foundation, the Participating
Institutions, the National Science Foundation, the U.S. Department
of Energy, the National Aeronautics and Space Administration, the
Japanese Monbukagakusho, the Max Planck Society, and the Higher
Education Funding Council for England.  This research has made use
of the NASA/IPAC Infrared Science Archive (IRSA), which is operated
by the Jet Propulsion Laboratory, California Institute of Technology,
under contract with the National Aeronautics and Space Administration.
This work is based in part on observations made with the {\it Spitzer
Space Telescope}, which is operated by the Jet Propulsion
Laboratory/California Institute of Technology, under a contract
with NASA.  This work is also based in part on observations made
with the NASA/ESA {\it Hubble Space Telescope}, obtained at the
Space Telescope Science Institute, which is operated by the Association
of Universities for Research in Astronomy, Inc., under NASA contract
NAS 5-26555.  RJA is supported by an appointment to the NASA
Postdoctoral Program at the Jet Propulsion Laboratory, administered
by Oak Ridge Associated Universities through a contract with NASA.

\copyright 2012.  All rights reserved.

\appendix

\section{Additional Spectroscopic Redshifts in the COSMOS Field}

The three slitmasks that we observed were designed to target
\wise-selected AGN candidates in the COSMOS field, though the low
source density of such sources allowed for additional spectroscopic
targets.  We primarily filled out the masks with IRAC-selected AGN
candidates, using the two-color criteria of \citet{Stern:05b}.
Given the interest and use of the COSMOS field by a broad community,
we include those additional sources here.

Table~\ref{table:appendix} presents the results for 26 COSMOS sources
for which we obtained redshifts; the six targeted sources are listed
in Table~\ref{table:keck}.  We include the quality (``Q'') of each
spectroscopic redshift.  Quality flag ``A'' signifies an unambiguous
redshift determination, typically relying upon multiple emission
or absorption features.  Quality flag ``B'' signifies a less certain
redshift determination, such as the robust detection of an isolated
emission line, but where the identification of the line is uncertain
\citep[\eg,][]{Stern:00d}.  Quality flag ``B'' might also be assigned
to a source with a robust redshift identification, but where some
uncertainty remains as to the astrometric identity of that spectroscopic
source.  We consider the quality ``B'' results likely to be correct,
but additional spectroscopy would be beneficial.  All of the
spectroscopic redshifts in Table~\ref{table:keck} are of quality
``A''.


\scriptsize
\begin{deluxetable}{ccccccl}
\tablecaption{Additional Results from Keck Observations.}
\tablehead{
\colhead{$I_{814}$} &
\colhead{R.A.} &
\colhead{Dec.} &
\colhead{$z$} &
\colhead{Q} &
\colhead{Slitmask(s)} &
\colhead{Notes}}
\startdata
17.32 & 10:00:11.83 & +02:26:23.1 & 0.440 & A & D[35] & [OII],[NeIII],H$\zeta$,H$\epsilon$,H$\delta$,H$\gamma$,H$\beta$,[OIII] \\
16.26 & 10:00:14.89 & +02:27:17.9 & 0.728 & A & D[38] & [OII],H$\beta$,[OIII] \\
19.30 & 10:00:22.79 & +02:25:30.6 & 0.349 & A & D[3]  & CaHK,H$\alpha$,[NII] \\
17.99 & 10:00:24.28 & +02:27:36.2 & 1.243 & B & D[39] & [OII] \\
17.14 & 10:00:24.51 & +02:26:18.1 & 1.129 & A & D[34] & [OII] \\
18.36 & 10:00:28.56 & +02:27:25.8 & 0.248 & A & D[4]  & CaH,H$\gamma$,H$\beta$,[OIII],H$\alpha$,[NII] \\
17.15 & 10:00:29.55 & +02:26:35.9 & 0.348 & B & D[36] & [OII],H$\beta$,[OIII],H$\alpha$,[NII] (could be serendip) \\
16.93 & 10:00:32.46 & +02:27:59.3 & 1.405 & B & D[41] & [OII] \\
16.80 & 10:00:33.23 & +02:27:59.3 & 0.981 & B & D[42] & [OII] \\
17.93 & 10:00:44.50 & +02:23:54.0 & 1.299 & B & D[17] & [OII],CaHK \\
18.43 & 10:00:50.15 & +02:26:18.5 & 3.730 & A & D[33] & QSO: Ly$\alpha$,CIV \\
18.56 & 10:00:50.58 & +02:23:29.3 & 3.093 & A & D[13] & QSO: Ly$\alpha$,CIV,CIII] \\
19.08 & 10:00:56.65 & +02:26:35.5 & 0.344 & A &B,D[0] & MgII absn,[OII],CaH,H$\gamma$,H$\beta$,[OIII],H$\alpha$,[NII] \\
17.55 & 10:00:58.07 & +02:26:16.8 & 0.425 & A & D[32] & [OII],H$\beta$,[OIII],H$\alpha$ \\
16.55 & 10:00:58.70 & +02:25:56.2 & 0.694 & A & D[28] & QSO: [NeV],[OII],H$\beta$,[OIII] \\
17.09 & 10:00:59.00 & +02:24:17.9 & 1.193 & B & B     & [OII] \\
19.15 & 10:00:59.81 & +02:24:30.7 & 0.541 & A & B     & [OII] \\
16.94 & 10:01:08.35 & +02:23:42.0 & 1.930 & A &B,D[15]& QSO: Ly$\alpha$,CIV,CIII],MgII \\
18.22 & 10:01:08.65 & +02:23:14.1 & 0.503 & A & D[12] & [OII],D4000 \\
16.52 & 10:01:13.93 & +02:25:48.1 & 0.373 & A & B     & QSO: broad MgII,[NeV],[OII],H$\alpha$ \\
17.18 & 10:01:14.68 & +02:24:49.5 & 1.656 & A & B     & LBG: CII,CIV,[OII] \\
16.36 & 10:01:17.00 & +02:27:31.2 & 0.518 & A & B     & H$\alpha$ \\
16.25 & 10:02:17.42 & +02:29:59.7 & 1.100 & A & A     & QSO: CIV,CIII]; odd broad lines at $\sim$ 1650 \AA \\
17.56 & 10:02:28.18 & +02:30:15.4 & 0.344 & B & A     & [OII] \\
15.23 & 10:02:29.89 & +02:32:25.1 & 0.431 & A & A     & AGN: [NeV],[OII],[NeIII],[OIII] \\
17.76 & 10:02:31.90 & +02:35:07.4 & 0.880 & A & A     & AGN: CIII],MgII \\
\enddata
\label{table:appendix}

\tablecomments{Q indicates the quality of the redshift (see text
for details).  Masks A and B were observed with LRIS.  Mask D was
observed with DEIMOS; the bracketed numbers indicate the DEIMOS
slitlet number.}

\end{deluxetable}


\clearpage
\end{document}